\documentclass[12pt]{article}
\usepackage{latexsym,epsfig,amssymb} 
\newcommand{\ft}[2]{{\textstyle\frac{#1}{#2}}}

\def\bfone{\relax{\rm 1\kern-.35em 1}}

\newcommand{\be}{\begin{equation}}
\newcommand{\ee}{\end{equation}}
\newcommand{\ben}{\begin{displaymath}}
\newcommand{\een}{\end{displaymath}}
\newcommand{\bea}{\begin{eqnarray}}
\newcommand{\eea}{\end{eqnarray}}

\newcommand{\non}{\nonumber\\}
\newcommand{\bean}{\begin{eqnarray*}}
\newcommand{\eean}{\end{eqnarray*}}

\makeatletter
\@addtoreset{equation}{section}
\makeatother

\begin{document}

\thispagestyle{empty}

\begin{flushright}\small 
ENSL-00325444
\end{flushright}

\bigskip
\bigskip

\vskip 10mm
\begin{center}
  {\LARGE {\bf Supergravities without an Action:\\[1.7ex]
   Gauging the Trombone}}
\end{center}


\bigskip
\medskip

\vskip 8mm

\begin{center}
{{\bf Arnaud Le Diffon and Henning Samtleben}\\[.4ex]
Universit\'e de Lyon, Laboratoire de Physique, ENS Lyon,\\ 
46 all\'ee d'Italie, F-69364 Lyon CEDEX 07, France \\
{\tt arnaud.le\_diffon, henning.samtleben @ens-lyon.fr}}
\vskip 4mm
\end{center}

\vskip2cm
\begin{center} {\bf Abstract } \end{center}
\begin{quotation}\noindent
We present a systematic account of supergravity theories
in which the global scaling symmetry is gauged.
This generalizes the standard gaugings of non-abelian
off-shell symmetries.
A particular feature of these theories is an additional
positive contribution to the effective cosmological constant.

As the scaling symmetry is an on-shell symmetry, the resulting
gaugings do no longer possess an action. 
We develop the algebraic framework for the maximal theories
in various dimensions and construct explicit solutions to the
algebraic consistency constraints ---
related to ``pure-spinor-like'' structures for the exceptional groups.
As an example, we explicitly work out the modified 
supersymmetry transformation
rules and equations of motion in three dimensions.
Finally, we speculate about the role of these theories
from the perspective of very extended Kac-Moody algebras.
\end{quotation}

\newpage
\setcounter{page}{1}

\tableofcontents

\bigskip
\bigskip


\section{Introduction}


Einstein's equations of general relativity
possess a well-known global symmetry under
conformal rescaling of the metric 
\bea
g_{\mu\nu}&\rightarrow& \Lambda^2 \,g_{\mu\nu}
\;,
\label{scaling}
\eea
with constant $\Lambda$. 
As the Einstein-Hilbert Lagrangian scales according to
${\cal L}_{\rm EH}\rightarrow\Lambda^{D-2} \,{\cal L}_{\rm EH}$,
this symmetry is off-shell realized only in $D=2$ space-time dimensions.
The global scaling symmetry extends to supergravity theories
in all dimensions, with gravitinos and matter fields scaling 
with different weights under (\ref{scaling}),
in particular, $p$-forms scale with weight $p$ and scalars fields are invariant.
In this context it is often referred to as a 
{\em trombone} symmetry
and plays an important role e.g.\ among the
spectrum-generating symmetries for 
the fundamental BPS solutions~\cite{Cremmer:1997xj}.

In addition, the maximal supergravity theories admit 
rather large global symmetry groups, given by
the exceptional groups ${\rm G}={\rm E}_{n(n)}$, which
(at least in odd dimensions) are genuine off-shell symmetries of the 
action.\footnote{The
notation ${\rm E}_{n(n)}$ refers to the split form of the exceptional 
group ${\rm E}_{n}$.}
These have first been revealed in four-dimensional maximal
supergravity~\cite{Cremmer:1979up} and we
will refer to them as duality (or Cremmer-Julia) symmetries,
see~\cite{Cremmer:1997ct} for a review.
In fact, the trombone symmetry~(\ref{scaling})
plays an important role in the realization of the Cremmer-Julia
groups~${\rm E}_{n(n)}$ in the various dimensions.
Recall that maximal supergravities are obtained by dimensional
reduction of the eleven-dimensional theory~\cite{Cremmer:1978km}
on an $n$-torus.
From the eleven-dimensional point of view there are 
two scaling symmetries inherited to the $D=(11\!-\!n)$-dimensional
theory. Apart from the trombone symmetry~(\ref{scaling})
of eleven-dimensional supergravity, a rescaling of the $n$-torus
\bea
y^i&\rightarrow& \alpha\, y^i \;,\quad i=1, \dots, n\;,\qquad
\mbox{for the coordinates~}y^i \mbox{~of the torus}
\;,
\label{torusscaling}
\eea
is part of the eleven-dimensional diffeomorphisms and
translates into a particular rescaling of the $D$-dimensional fields.
From the $D$-dimensional point of view, it is however more natural
to consider particular combinations of the two scaling 
symmetries~(\ref{scaling}), (\ref{torusscaling}):
Choosing $\alpha=\Lambda$ defines a scaling which leaves the scalars
of the $D$-dimensional theory invariant --- this is the $D$-dimensional
trombone symmetry~(\ref{scaling}). On the other hand, 
choosing $\alpha=\Lambda^{9/n}$ 
defines a scaling which leaves the $D$-dimensional metric 
(in the Einstein frame)
invariant;\footnote{In the reduction to $D=2$ dimensions ($n=9$),
this scaling degenerates as a sign of the fact that
in two dimensions the theory cannot be cast into the Einstein frame.}
this symmetry is part of the Cremmer-Julia group embedded as
${\rm GL}(1)\subset {\rm GL}(n)\subset {\rm E}_{n(n)}\,$.
This shows how higher-dimensional trombone symmetries naturally
merge with the lower-dimensional duality groups.

It is well known that certain subgroups of the global
Cremmer-Julia symmetry groups ${\rm E}_{n(n)}$
may be promoted to local symmetries while preserving all 
supersymmetries~\cite{deWit:1982ig,Gunaydin:1985cu,Pernici:1984xx}.
The resulting {gauged supergravities} exhibit non-abelian gauge
groups, additional couplings and in particular a scalar potential.
The construction of these theories can be 
systematically performed using the group-theoretical framework
of~\cite{Nicolai:2000sc,deWit:2002vt,deWit:2004nw,deWit:2005hv}
which allows to characterize the various gaugings in terms of
a single tensorial object, the {\em embedding tensor} $\Theta$, subject to a number
of algebraic constraints that encode the consistency of the theory.
In view of the close relation of the Cremmer-Julia groups ${\rm E}_{n(n)}$ and the
trombone symmetry~(\ref{scaling}) in dimensional reduction, 
it seems natural to also consider the possible gaugings of the 
trombone symmetry.
This is what we are going to address in this paper.

Certain theories with local trombone symmetry
have already appeared in the literature.
A straightforward way to obtain such theories is by performing
a standard Scherk-Schwarz reduction~\cite{Scherk:1979zr}
twisting the fields with the higher-dimensional 
on-shell symmetry~(\ref{scaling}).
Applying this to the circle reduction of eleven-dimensional
supergravity gives rise to a one-parameter 
deformation of the ten-dimensional IIA theory with maximal 
supersymmetry~\cite{Howe:1997qt,Lavrinenko:1997qa}
which is different from Romans' massive supergravity~\cite{Romans:1985tz}.
In particular, this theory does not have an action and admits a de Sitter vacuum.
According to the discussion above, from the ten-dimensional point of view 
this theory corresponds to the gauging of a linear combination of the $D=10$ trombone 
symmetry and the off-shell ${\rm GL}(1)$ symmetry.
It has been further studied 
in~\cite{Chamblin:1999ea,Chamblin:2001dx,Chamblin:2001jj,Gheerardyn:2002wp}.
Other examples of such theories have been obtained in lower dimensions by
studying analogous generalized Scherk-Schwarz 
reductions to nine and to six dimensions~\cite{Bergshoeff:2002nv,Kerimo:2003am,Kerimo:2004md}.

In this paper we will set up a systematic framework 
for the classification and construction of these theories.
We follow the group-theoretical approach 
of~\cite{Nicolai:2000sc,deWit:2002vt,deWit:2004nw,deWit:2005hv},
in which theories with a local trombone symmetry~(\ref{scaling})
simply correspond to the introduction of additional components $\theta_M$ in
the embedding tensor.
This allows to straightforwardly derive consistency conditions on
such gaugings and to construct explicit examples exploiting the
structure of the underlying symmetry groups.
As it turns out, the additional components in the embedding tensor 
generically induce a simultaneous 
gauging of the trombone symmetry~(\ref{scaling}) and a subgroup 
of the duality group~${\rm G}$.

\begin{table}[tb]
\begin{center}
\begin{tabular}{r||c|c|c||c|c|}
$D$ & ${\rm G}$ & ${\cal R}_{\rm adj}$ & ${\cal R}_{\rm v}$ & $\Theta$  &$\theta$    \\
\hline
7   & ${\rm SL}(5)$   &  ${\bf 24}$ & ${\bf 10}'$ & ${\bf 15}+{\bf 40}'$ & ${\bf 10}$\\
6   &  ${\rm SO}(5,5)$  &  ${\bf 45}$ & ${\bf 16}_c$ & ${\bf 144}_c$ & ${\bf 16}_s$\\
5   & E$_{6(6)}$   &  ${\bf 78}$ & ${\bf 27}'$ & ${\bf 351}'$ & ${\bf 27}$\\
4   & E$_{7(7)}$  &  ${\bf 133}$ & ${\bf 56}$ & ${\bf 912}$ & ${\bf 56}$ \\
3   & E$_{8(8)}$  & ${\bf 248}$ & ${\bf 248}$ & ${\bf 1}\!+\!{\bf 3875}$ & ${\bf 248}$  \\
\cline{5-6}
2   & E$_{9(9)}$  & $\;\;{\bf{\cal R}}_{\rm adj}$ & ${\bf \Lambda}_1$ & 
\multicolumn{2}{c}{$\;\;\;\;\;{\bf \Lambda}_{1*}$} 
\end{tabular}
\end{center}
\caption{\small
The embedding tensor in maximal supergravity in various dimensions. 
${\cal R}_{\rm adj}$ and ${\cal R}_{\rm v}$ denote the adjoint representation of
the global symmetry group ${\rm G}$ and the representation
in which the vector fields transform, respectively.
The tensors $\Theta$ and $\theta$ denote the components of the
embedding tensor, the latter also induce a gauging of the trombone 
symmetry~(\ref{scaling}).
}\label{tabTT}
\end{table}

In table~\ref{tabTT} we have collected the representations
in which the embedding tensor transforms for the maximal supergravities in various dimensions.
The standard gaugings are described by a tensor $\Theta$ 
inducing gauge groups that are subgroups of the duality group ${\rm G}$ and do not
include the trombone symmetry~(\ref{scaling}).
This tensor transforms in a particular subrepresentation of the tensor product
${\cal R}_{\rm adj}\otimes {\cal R}_{{\rm v}*}$ of the adjoint representation
of ${\rm G}$ and the representation dual to the vector fields of the theory.
The corresponding theories have been constructed 
in~\cite{Nicolai:2000sc,deWit:2004nw,Samtleben:2005bp,deWit:2007mt,Samtleben:2007an,Bergshoeff:2007ef}. 
The theories we will construct in this paper allow for additional components
in the embedding tensor, combining into a vector $\theta$ which transforms in 
the representation ${\cal R}_{{\rm v}*}$.
Note that in two dimensions the two objects $\Theta$ and $\theta$ coincide
(in fact, this observation triggered the present investigation).
As a consequence, in two dimensions gaugings generically include a local
trombone symmetry~(\ref{scaling}), in accordance with the fact that 
in $D=2$ dimensions this symmetry becomes off-shell ---
more precisely it builds the central extension of the affine
global symmetry group ${\rm E}_{9(9)}$~\cite{Julia:1981wc}.
Moreover, this indicates that the new theories we present in this paper
are particularly interesting from the unifying point of view of the extended 
Kac-Moody algebras ${\rm E}_{10}$~\cite{Damour:2002cu} and 
${\rm E}_{11}$~\cite{West:2001as}, conjectured to underlie 
eleven-dimensional supergravity and its compactifications.
We will come back to this in the conclusions.

The rest of this paper is organized as follows.
In section~\ref{sec:gauging} we set up the general
formalism in order to describe a theory with
local scaling symmetry~(\ref{scaling}).
In the gravity sector this introduces new minimal couplings 
between the metric and a vector field which 
modify the Einstein equations. 
In the full theory we describe the simultaneous gauging
of a subgroup of the duality group ${\rm G}$
and the scaling symmetry~(\ref{scaling}) by an embedding
tensor $\widehat{\Theta}$ which completely encodes the theory.
With respect to the standard constructions, the possibility of a local scaling symmetry
translates into a set of additional components $\theta_M$ of the embedding tensor.
We derive the quadratic constraints on the embedding
tensor which encode consistency of the gaugings.

In section~\ref{sec:algebraic} we work out the details of the construction for
all maximal supergravities in dimensions $6\ge D\ge3$.
The analysis relies on the particular properties of the 
global symmetry groups ${\rm SO}(5,5)$, ${\rm E}_{6(6)}$, 
${\rm E}_{7(7)}$, and ${\rm E}_{8(8)}$ of these theories.
In particular, we investigate the class of gaugings which is exclusively
triggered by the new components $\theta_M$ of the embedding tensor.
In this case, the quadratic consistency constraints reduce to a simple set
of equations that for ${\rm SO}(5,5)$ reduce to the well-known 
``pure-spinor'' condition 
and to its higher-rank analogues in the other dimensions. We present the
explicit solution of these quadratic constraints.

Subsequently, in section~\ref{sec:supersymmetry} we analyze the
compatibility of the gaugings with supersymmetry. For the 
case of the three-dimensional theory we derive the full set of deformed
equations of motion and show closure of the supersymmetry algebra.
We find that a particular effect of the theories with local scaling symmetry
is a positive contribution to the effective cosmological constant.
We close the paper with some speculations on the possible role
of these new theories in the unifying framework of the
extended Kac-Moody algebras ${\rm E}_{10}$ and ${\rm E}_{11}$.


\section{Gauging the scaling symmetry}
\label{sec:gauging}


In supergravity theories, the trombone symmetry~(\ref{scaling})
extends to the full bosonic field content: the metric 
and the antisymmetric $p$-forms infinitesimally scale as
\bea
\delta g_{\mu\nu} &=& 2\lambda\,g_{\mu\nu} \;,\qquad
\delta A_{\mu_1 \dots \mu_p} ~=~ p\lambda\,A_{\mu_1 \dots \mu_p}
\;,
\label{Rbos}
\eea
respectively, with a constant parameter $\lambda$,
while scalar fields remain invariant.
The fermionic fields on the other hand transform as
\bea
\delta \psi_\mu &=& \ft12\lambda\,\psi_\mu  \;,\qquad
\delta\chi ~=~ -\ft12\lambda\,\chi 
\;,
\label{Rferm}
\eea
for gravitinos $\psi_\mu$ and spin-1/2 fermions $\chi$, 
respectively. It is easy to check
that under this symmetry all kinetic terms of the Lagrangian
scale homogeneously as
\bea
\delta {\cal L}_{\rm kin} &=& (D\!-\!2)\,\lambda\,{\cal L}_{\rm kin}
\;.
\eea
It is a non-trivial property of supergravity theories that also 
all interaction terms scale with the same weight.
In particular, this restricts the topological terms to
two-derivative terms.

In the following we will consider gaugings of supergravity in which the
trombone symmetry (\ref{Rbos}), (\ref{Rferm}) becomes a local symmetry. 
Following the standard procedure, this is achieved by introducing
covariant derivatives
\bea
D_\mu &\equiv& \partial_\mu -  {\cal A}_\mu\,t_0
\;,
\label{covt0}
\eea
where $t_0$ denotes the generator of the trombone symmetry.
However, this cannot be the full answer.
Consistency implies that the vector field ${\cal A}_\mu$ 
itself must not be charged under the symmetry it is gauging.
This shows already that the local gauge symmetry cannot 
simply be the scaling symmetry (\ref{Rbos}), (\ref{Rferm}) under which
all vector fields are charged.
Rather, gauging of the scaling symmetry must be accompanied by
a gauging of other generators of the global symmetry group ${\rm G}$ 
of the theory --- which however is invisible in the gravity sector.
Indeed, this is what we will find in the following.


\subsection{Gravity sector}


To begin with, we will study the gravity sector with local scaling symmetry~(\ref{Rbos}),
i.e.\ introduce covariant derivatives~(\ref{covt0}) in Einstein's equations.
Since eventually we are interested in supergravity, we use the formulation
in terms of the vielbein $e_\mu{}^a$ with curved indices $\mu, \nu, \dots$ and
flat indices $a, b, \dots$.
As a first step, the standard spin connection $\omega_{\mu}{}^{ab}$ defined by
\bea
0&\stackrel!{\equiv}&
\nabla(\omega)_{[\mu} e_{\nu]}{}^a
~\equiv~
\partial_{[\mu} e_{\nu]}{}^a ~+~ \omega_{[\mu}{}^{ab}\,e_{\nu]\,b} \;, 
\eea
is replaced by a covariantized object $\widehat\omega_{\mu}{}^{ab}$
defined by
\bea
0&\stackrel!{\equiv}&
D_{[\mu} e_{\nu]}{}^a ~+~ \widehat\omega_{[\mu}{}^{ab}\,e_{\nu]\,b} \;, 
\eea
with the covariant derivative $D_\mu$ from (\ref{covt0}). Explicitly, this yields
\bea
\widehat\omega_{\mu}{}^{ab} &=&\omega_{\mu}{}^{ab}
-2 \,e_\mu{}^{[a}\,\,e^{ b]\, \nu}{\cal A}_{\nu}
\;,
\eea
for the modified spin connection which is uncharged under the 
scaling symmetry.
It is important to note that this covariantization of the spin connection may
equivalently be interpreted as the adding of a torsion trace term
$T_{\mu\nu}^a= 2 \,{\cal A}_{[\mu}\,e_{\nu]}{}^a$.\footnote{The equivalence
breaks down once we consider additional matter in the theory.}
Likewise, we define the covariantized Riemann tensor as
\bea
\widehat{{\cal R}}_{\mu \nu}{}^{ab} &\equiv&
2\,\partial_{[\mu}\,\widehat\omega_{\nu]}{}^{ab} + 
2\, \widehat\omega_{[\mu}{}^{ac}\:\widehat\omega_{\nu]c}{}^b
 \nonumber\\[1ex]
&=& {R}_{\mu \nu}{}^{ ab} 
+ 4 \, e_{[\mu}{}^{[a} \nabla(\omega)_{\nu]}{\cal A}^{b]} + 4 \, e_{[\mu}{}^{[a} {\cal A}_{\nu]} {\cal A}^{b]} 
- 2 \, e_{[\mu}{}^{a} e_{\nu]}{}^{b} {\cal A}_{\lambda} {\cal A}^{\lambda}
\;.
\label{Riemann}
\eea
By construction it is invariant under gauge transformations
\bea
\delta\,e_\mu{}^a &=& \lambda(x)\,e_\mu{}^a\;,\qquad
\delta{\cal A}_\mu ~=~ \partial_\mu\lambda(x)
\;.
\label{gaugeg}
\eea
The covariantized Riemann tensor~(\ref{Riemann})
no longer possesses the symmetries of the standard Riemann tensor:
the first Bianchi identity is modified to
\bea
\widehat{{\cal R}}_{[\mu \nu\rho]}{}^{a} 
&=&
 -  {\cal F}_{[\mu\nu}\,e_{\rho]}{}^{a}
\;,
\label{Bianchi}
\eea
with the abelian field strength ${\cal F}_{\mu\nu}=2\,\partial_{[\mu}{\cal A}_{\nu]}$.
For later supergravity calculations we also note the relation
\bea
\widehat{{\cal R}}_{\mu\nu}{}^{ab}\,\gamma^{\rho\mu\nu}\,\gamma_{ab}
&=& 
4\,\Big(\widehat{{\cal R}}^{(\rho\mu)}-\ft12g^{\rho\mu}\,\widehat{{\cal R}}\Big)\,\gamma_\mu
-2\,(D-3)\gamma^{\rho\mu\nu}\,{\cal F}_{\mu\nu}-2\,(D-2)\,{\cal F}^{\rho\mu}\gamma_\mu
\;,
\non
\label{sugraBianchi}
\eea
with $\gamma$-matrices in $D$ space-time dimensions,
and the Ricci tensor $\widehat{{\cal R}}_{\mu\nu}=\widehat{{\cal R}}_{\mu\rho\nu}{}^b\,e_b{}^\rho$\,
and Ricci scalar $\widehat{{\cal R}}=g^{\mu\nu}\widehat{{\cal R}}_{\mu\nu}$\,.
Explicitly, the latter are given by\footnote{Our notation here is such that $\nabla_\mu$
in these equations refers to the covariant derivative $\nabla(\Gamma)_\mu$ in presence
of the standard (non-covariantized) Christoffel symbols $\Gamma^\lambda_{\mu\nu}$.}
\bea
\widehat{{\cal R}}_{(\mu \nu )} &=&
 {R}_{\mu \nu} +
 (D-2) \, \Big(\nabla_{(\mu} {\cal A}_{\nu)} 
 + {\cal A}_{\mu} {\cal A}_{\nu}\Big)
 + g_{\mu \nu}\,\Big(
  \nabla^\lambda {\cal A}_{\lambda} 
  -(D-2)\,A^\lambda A_\lambda \Big)
 \;,
 \nonumber\\[1ex]
\widehat{{\cal R}} &=& {R} + 2\,(D-1) \, g^{\lambda \rho} \nabla_{\lambda} {\cal A}_{\rho} -(D-1)(D-2) \, {\cal A}_{\lambda} {\cal A}^{\lambda} 
\;.
\eea
The covariantized Einstein equations (in absence of matter)
thus are 
\bea
\widehat {{\cal R}}_{(\mu \nu)} - \ft{1}{2} \, \widehat{{\cal R}}\, g_{\mu \nu} &=& 0
\;, 
\label{Einstein0mod}
\eea
and manifestly invariant under~(\ref{gaugeg}).
In the theories we will consider, additional matter will always be present,
in particular a gauge field sector which includes the vector field~${\cal A}_\mu$,
such that the right-hand side of the Einstein equations will
be non-vanishing, see e.g.~equation~(\ref{eqmEinstein}) below.
One may verify that equations~(\ref{Einstein0mod})
do no longer descend from a standard action which is expected
since we have gauged a symmetry that was not off-shell realized.

In the following we will extend the gauging to the remaining 
matter fields of supergravity. In particular, we need to identify 
among the supergravity gauge fields the vector
field ${\cal A}_\mu$ used in all covariant derivatives.


\subsection{Embedding tensor}


Extending the gauging to the full theory is conveniently
described by resorting to the group-theoretical formalism
developed in~\cite{Nicolai:2000sc,deWit:2002vt,deWit:2004nw,deWit:2005hv}.
As we have discussed in the introduction, the full global symmetry group of the ungauged
theory is given by the direct product $\mathbb{R}^+\times{\rm G}$
where the first factor describes the scaling (\ref{Rbos}), (\ref{Rferm})
and the second factor is the standard duality group.
We will denote the total set of generators by
$\{ t_{\hat\alpha} \} = \{t_0, t_\alpha \}$, $\hat{\alpha}=0, \dots, {\rm dim~G}$,
where $t_0$ denotes the generator of $\mathbb{R}^+$ and the
$t_\alpha$ denote the generators of ${\rm G}$.
The latter satisfy commutation relations
\bea
{}[t_\alpha,t_\beta] &=& f_{\alpha\beta}{}^\gamma\,t_\gamma\;.
\eea
The vector fields $A_\mu^M$ in the ungauged theory 
transform in some representation ${\cal R}_{\rm v}$ of ${\rm G}$ labeled by 
$M=1, \dots, {\rm dim~}{\cal R}_{\rm v}$,
and carry charge $+1$ under $\mathbb{R}^+$ according to~(\ref{Rbos}).

A general gauging is defined by introducing covariant derivatives
\bea
D_\mu &\equiv& \partial_\mu - 
g A^M_\mu\, \widehat\Theta_M{}^{\hat{\alpha}}\,t_{\hat\alpha}
~=~
\partial_\mu - 
g A^M_\mu\, \widehat\Theta_M{}^{0}\,t_{0} - 
g A^M_\mu\, \widehat\Theta_M{}^{{\alpha}}\,t_{\alpha}
\;,
\label{covariant}
\eea
in terms of an embedding tensor $\widehat\Theta_M{}^{\hat{\alpha}}$
which describes the embedding of the gauge group generators
$X_M\equiv \widehat\Theta_M{}^{\hat{\alpha}} t_{\hat\alpha}$ into the 
symmetry group of the ungauged theory.
In addition, we have introduced the gauge coupling constant $g$. 
It can in principle be absorbed into the embedding tensor.

According to its coupling the embedding tensor carries charge $-1$ under $\mathbb{R}^+$.
Its component $\theta_M\equiv \widehat\Theta_M{}^0$ 
transforms under ${\rm G}$ in the representation ${\cal R}_{\rm v*}$ 
dual to ${\cal R}_{\rm v}$. It selects the vector field that gauges the $\mathbb{R}^+$-symmetry.
Comparing (\ref{covariant}) to (\ref{covt0}) we identify
\bea
{\cal A}_\mu &=& g\,\theta_M\,A^M_\mu\;.
\eea
The remaining part of the connection (\ref{covariant}) involves the 
generators of the duality group ${\rm G}$ and is thus
invisible in the gravity sector discussed above.
It is defined by the component $\widehat\Theta_M{}^\alpha$ of the embedding tensor
which a priori transforms in the tensor product
\bea
{\cal R}_{\rm v*} \otimes {\cal R}_{\rm adj} &=& 
{\cal R}_{\rm v*} \oplus~ \dots \;,
\label{tensor}
\eea
with ${\cal R}_{\rm adj}$ denoting the adjoint representation of ${\rm G}$.
For gaugings that do not involve the scaling symmetry $\mathbb{R}^+$,
it is known that supersymmetry restricts the allowed choices for $\widehat\Theta_M{}^\alpha$
to only very few of the irreducible representations on the r.h.s.\ of 
(\ref{tensor}), see e.g.~\cite{deWit:2002vt}.
In particular, in the maximal theories (for $3\le D\le 7$), 
the ``trace part'' ${\cal R}_{\rm v*}$ in this tensor product
is always forbidden.\footnote{
In the half-maximal theories, the representation ${\cal R}_{\rm v*}$ 
appears with multiplicity 2 on the r.h.s.\ of (\ref{tensor}) and supersymmetry
implies a linear relation between these two representations~\cite{Schon:2006kz}.
The same happens for the maximal theories in $D=8, 9$, where the 
group ${\rm G}$ is no longer simple~\cite{Bergshoeff:2002nv}.}
In $D=2$ space-time dimensions on the other hand, the picture is quite the opposite:
the gaugings are precisely parametrized by the ``trace part''  
${\cal R}_{\rm v*}$  on the r.h.s.\ of  (\ref{tensor}),
i.e.\
gaugings are described by a vector~$\theta_M$ transforming in the 
basic representation of the infinite-dimensional affine algebra 
${\rm E}_{9(9)}$~\cite{Samtleben:2007an}.
From a unifying point of view of the gauged supergravities this is somewhat
unsatisfactory; upon dimensional reduction the algebraic structures in higher
dimensions are typically embedded into the lower-dimensional structures
described by higher-rank Kac-Moody algebras. 
We will see that precisely the theories that involve a
gauging of the trombone symmetry close this gap and allow
for a non-vanishing ${\cal R}_{\rm v*}$ in (\ref{tensor}) also in dimensions $D>2$.
Recall that only in $D=2$ dimensions, the trombone symmetry
is part of the off-shell symmetries of the action and 
shows up as the central extension
of the affine algebra ${\rm E}_{9(9)}$.

Our general ansatz for the embedding tensor and thus for the
connection in (\ref{covariant}) is the following:
\bea
\widehat\Theta_M{}^0 &=& \theta_M\;,\qquad
\widehat\Theta_M{}^\alpha ~=~ 
\Theta_M{}^\alpha + \zeta\, \theta_N\,(t^\alpha)_M{}^N
\;,
\label{ansatz}
\eea
where we have split $\widehat\Theta_M{}^\alpha$ into a ``traceless''  part satisfying
$\Theta_M{}^\alpha\,(t_\alpha)_N{}^M=0$
and its ``trace part'', corresponding to the representation ${\cal R}_{\rm v*}$
on the r.h.s.\ of~(\ref{tensor}).\footnote{
Here and in the following, we raise and lower adjoint indices
with the invariant metric $\kappa_{\alpha\beta}\equiv {\rm Tr}[t_{\alpha}t_{\beta}]$
which is related to the Cartan-Killing form $\eta_{\alpha\beta}$
as 
$\kappa_{\alpha\beta}=C_{\rm v}({\rm dim}{\cal R}_{\rm v})/ ({\rm dim}{\cal R}_{\rm adj})\,\eta_{\alpha\beta}$ with the Casimir operator $C_{\rm v}$ in the vector field representation.
}
As this second term is forbidden in the standard gaugings, 
it is natural to assume that it comes
proportional to the same vector $\theta_M$ 
that triggers the gauging of the 
trombone symmetry via $\widehat{\Theta}_M{}^0$.
We will explicitly verify this assumption.
All that remains in this ansatz is to determine the proportionality constant $\zeta$,
which must be done case by case, i.e.\ in dependence of the space-time dimension
and the number of supercharges.

To this end, we recall that a generic gauging
introduces non-trivial couplings between vector fields and the antisymmetric
two-form tensors.
E.g.\ the non-abelian field strength of the vector fields receives
corrections of the St\"uckelberg type~\cite{deWit:2004nw,deWit:2005hv}
\bea
{\cal F}^M_{\mu\nu} &\longrightarrow&
{\cal H}^M_{\mu\nu} ~\equiv~
{\cal F}^M_{\mu\nu} + g\,Z^M{}_{PQ}\,B_{\mu\nu}^{PQ}
\;,
\label{Hgen}
\eea
with two-forms $B_{\mu\nu}^{PQ}$ and the intertwining tensor given by
\bea
Z^{M}{}_{PQ}&\equiv&(t_{\hat\alpha})_{(P}{}^M\, \widehat\Theta_{Q)}{}^{\hat\alpha}\,
\;.
\label{defZ}
\eea
In particular, this tensor encodes the field content of two-forms in the theory:
as in general in its indices $PQ$ it does not project onto the full symmetric
tensor product ${\cal R}_{\rm v}\otimes{\cal R}_{\rm v}$, 
but rather satisfies some non-trivial projection
\bea
Z^{M}{}_{PQ} &=& Z^{M}{}_{RS}\,{\mathbb P}^{RS}{}_{PQ}
\;,
\label{ZP}
\eea
also the two-forms will only appear under projection with ${\mathbb P}$,
see~\cite{deWit:2005hv,deWit:2008ta} for details.
We will take this projection as a guide to determine the
constant $\zeta$ in (\ref{ansatz}): the projector ${\mathbb P}$ in~(\ref{ZP})
and thus the two-form field content
should be the same in presence and in absence of an ${\mathbb{R}}^+$-gauging.
In particular, this is necessary because the two-form field content of the theory is 
fixed by supersymmetry.
We will further illustrate this argument and the calculation
in section~\ref{sec:algebraic}
in several examples.
Later in section~\ref{sec:supersymmetry} we also explicitly
confirm consistency
of the ansatz~(\ref{ansatz}) with supersymmetry.


\subsection{Quadratic constraints}


Before applying the above construction to various theories, let us
collect a few general formulae and relations.
With the ansatz (\ref{ansatz}) for the embedding tensor, 
the generators of the gauge group
evaluated in the vector field representation are given by
\bea
(X_M)_N{}^K \;\equiv\;
\widehat\Theta_M{}^{\hat{\alpha}}(t_{\hat{\alpha}})_N{}^K 
\;=\;
\Theta_M{}^\alpha (t_\alpha)_N{}^K
+ \Big( \zeta(t^\alpha)_M{}^P(t_\alpha)_N{}^K-\delta_M^P\delta_N^K \Big)\,
\theta_P
\;.
\label{defX}
\eea
In particular, this gives an explicit expression for the intertwining tensor
$Z$ from (\ref{defZ})
\bea
Z^{M}{}_{PQ}&=&
(t_\alpha)_{(P}{}^M\Theta_{Q)}{}^\alpha 
+ \Big( \zeta(t^\alpha)_P{}^{(K}(t_\alpha)_Q{}^{L)}-\delta_P^{(K}\delta_Q^{L)} \Big)\,
\theta_L
\;,
\label{evalZ}
\eea
from which we will determine the values of $\zeta$ in the examples below.
In every particular theory, supersymmetry will constrain the possible form
of $\Theta_M{}^\alpha$, this gives rise to the so-called linear representation
constraints on the embedding tensor.
In addition, consistency requires the components $\theta_M$, $\Theta_M{}^\alpha$
of the embedding tensor to satisfy a set of rather generic quadratic constraints.
These express the fact that the embedding tensor itself is invariant under the action
of the gauge group.
Evaluating this condition for the different components gives rise to the equations
\bea
0~\stackrel!{\equiv}~
\delta_M\,\theta_N&=&
\widehat\Theta_M{}^{\hat\alpha}\,\delta_{\hat\alpha}\,\theta_N
\nonumber\\[.4ex]
&=&
 (t_{\gamma})_{N}{}^{ Q}\, \Theta_M{}^\gamma\,\theta_{ Q}
+
\Big( \zeta\, (t^\gamma)_M{}^K (t_{\gamma})_{N}{}^{ L}
-\delta^K_M\delta^L_N \Big)\,
\theta_{ K} \theta_{ L}
\;,
\label{Q1}
\eea
and
\bea
0~\stackrel!{\equiv}~
\delta_P\,\Theta_M{}^\alpha 
\!&=&\!
\Theta_{ P}{}^{\beta} (t_{\beta})_{M}{}^{ N}
   \Theta_{ N}{}^\alpha +
    \Theta_{ P}{}^\beta f_{\beta\gamma}{}^\alpha 
    \Theta_{ M}{}^\gamma  
        \nonumber\\
    &&{}\!\!\!\!
    + \Big(
    \zeta\,\delta_\beta^\alpha\, (t^\gamma)_P{}^Q
     (t_{\gamma})_{M}{}^{ N}
     -\zeta\, f_{\beta\gamma}{}^\alpha\,\delta_{M}^N\, (t^\gamma)_{P}{}^Q 
          -\delta^Q_P\,\delta^N_M\,\delta^\alpha_\beta
    \Big) 
    \, \theta_Q \Theta_N{}^\beta
 \;.
 \nonumber\\
 \label{Q2}
\eea
Together, they guarantee in particular that the gauge group generators
(\ref{defX}) satisfy
\bea
{}[X_M,X_N] &=& -(X_{M})_{N}{}^K\,X_K \;,
\eea
i.e.\ these constraints ensure closure of the gauge algebra.
We will show in the following for various theories
that every set of $\theta_M$, $\Theta_M{}^\alpha$
that satisfies the given linear representation constraints and 
the quadratic relations (\ref{Q1}), (\ref{Q2}) defines a consistent
gauging which in case $\theta_M\not=0$ involves a gauging
of the trombone symmetry.

Let us finally note that the quadratic constraints~(\ref{Q1}), (\ref{Q2})
in particular imply the relations
\bea
\theta_M\,Z^M{}_{PQ} &=& 0 ~=~
\widehat{\Theta}_M{}^\alpha\,Z^M{}_{PQ}
\;,
\label{QQZ}
\eea
i.e.\ orthogonality between the embedding tensor and the 
intertwining tensor $Z$ from (\ref{evalZ}). This plays an important
role in the hierarchy of antisymmetric $p$-forms~\cite{deWit:2005hv,deWit:2008ta}.


\section{Algebraic structure in various dimensions}

\label{sec:algebraic}


In this section, we work out the above construction for the
maximal supergravities in dimensions $6\ge D\ge3$, for which the
global symmetry groups are given by ${\rm SO}(5,5)$, ${\rm E}_{6(6)}$, 
${\rm E}_{7(7)}$, and ${\rm E}_{8(8)}$, respectively.
We determine for the various cases 
the value of the parameter~$\zeta$ in (\ref{ansatz}),
which completely fixes the algebraic structure.
Furthermore, we evaluate the quadratic constraints (\ref{Q1}), (\ref{Q2})
and show that they admit non-trivial solutions.
As a result we obtain the full set of consistency constraints
for gaugings that involve a local trombone symmetry~(\ref{scaling}).

We will discuss in most detail the case $D=6$ in which the symmetry group
is the smallest and accordingly the algebraic structures are the 
simplest ones. Subsequently, we report
the results in lower dimensions which are obtained in complete 
analogy with slightly bigger computational effort.


\subsection{$D=6$}
\label{subsec:D6}


The ungauged theory in $D=6$ dimensions was constructed in~\cite{Tanii:1984zk},
its general gaugings were given in~\cite{Bergshoeff:2007ef},
to which we refer for details of the structure.
The global symmetry group of the ungauged theory
is the orthogonal group ${\rm SO}(5,5)$. 
Vector and two-form fields of this theory transform in the 
${\bf 16}_c$ and ${\bf 10}$ representations, respectively,
with the generators $t_\alpha=t_{[ij]}$ given by
\bea
(t_{ij})_M{}^N&=&(\gamma_{{ij}})_M{}^N\;,\qquad
(t_{ij})_k{}^{l}~=~ 4\,\eta_{k[i}\,\delta_{j]}^{\,l}\;,
\eea
respectively. Here, $i, j, \dots =1, \dots, 10$, and $M, N =1, \dots, 16$,
label the vector and the spinor representation of ${\rm SO}(5,5)$, respectively.
The tensors $\eta_{ij}$ and 
$(\gamma_{{ij}})_M{}^N$ denote the invariant form and the gamma matrices 
of ${\rm SO}(5,5)$, respectively. 
We use the former to raise and lower vector indices.
A non-trivial relation among the generators
that we will exploit in the following, is\footnote{
As mentioned above, adjoint indices are raised and lowered with
the invariant form $\kappa_{\alpha\beta}\equiv {\rm Tr}[t_{\alpha}t_{\beta}]$ which
here is given by $\kappa_{ij,kl}=-32\,\delta_{i[k}\delta_{l]j}$\,.}
\bea
(t^\alpha)_M{}^{K} (t_\alpha)_N{}^{L} &=&
-\ft1{32}\,(\gamma^{ij})_M{}^{K}\,(\gamma_{ij})_N{}^{L} 
~=~
\ft1{16}\,\delta_M^K\delta_N^L+\ft1{4}\delta_M^L\delta_N^K
-\ft18(\gamma^i)_{MN}(\gamma_i)^{KL}
\;,
\nonumber\\
\label{relD6}
\eea
which can be proven by further contraction with gamma matrices.

The embedding tensor $\widehat\Theta_M{}^\alpha$ describing the 
generators within the duality group ${\rm SO}(5,5)$ a priori lives
in the tensor product
\bea
{\bf 16}_s \otimes {\bf 45} &=& 
{\bf 16}_s \oplus {\bf {144}}_c \oplus {\bf 560_s}
\;,
\label{dec6}
\eea
where ${\bf 45}$ is the adjoint of ${\rm SO}(5,5)$.
In absence of the $\mathbb{R}^+$-gauging, supersymmetry
restricts the embedding tensor $\Theta_M{}^\alpha$ to the irreducible
${\bf {144}}_c$ representation in this decomposition~\cite{deWit:2002vt,Bergshoeff:2007ef}.
I.e.\ it can be parametrized in terms of a gamma-traceless vector-spinor~$\theta^{Mi}$ as 
\bea
\Theta_M{}^{ij}&=&-\theta^{N[i}\,\gamma^{j]}{}_{NM} \;,
\qquad
\mbox{with}\quad
\gamma_{i\,MN}\,\theta^{Ni} \equiv0
\;.
\eea
In this case, the intertwining tensor (\ref{defZ})
is given by
\bea
Z^{M}{}_{PQ} &=&
-\theta^{Li}\,\gamma^{j}{}_{L(P}\,(\gamma_{ij})_{Q)}{}^M
~=~ -(\gamma_i)_{PQ}\,\theta^{Mi}
\;,
\label{Z6}
\eea
where we have used the properties of the 
${\rm SO}(5,5)$ gamma matrices.
The form of~(\ref{Z6}) shows that in indices $(PQ)$, this tensor 
projects onto a subrepresentation
\bea
({\bf 16}_c \otimes {\bf 16}_c)_{\rm sym} &\longrightarrow& {\bf 10}
\;,
\label{proj6}
\eea
within the full symmetric tensor product.
According to the general discussion above, this must reproduce the
field content of the two-forms of the theory.
E.g.\ the general coupling (\ref{Hgen}) reduces to
\bea
{\cal H}_{\mu\nu}^M &=&
{\cal F}_{\mu\nu}^M - g \theta^{Mi} B_{\mu\nu\,i}
\;,
\eea
with two-forms $B_{\mu\nu\,i}\equiv (\gamma_{i})_{PQ} B_{\mu\nu}^{PQ}$ 
transforming in the ${\bf 10}$.
Indeed, this precisely coincides with the field content of
the ungauged theory.
Remarkably, this gives a purely
bosonic justification of the restriction of the embedding tensor
within~(\ref{dec6}).
Any other component in $\Theta_M{}^\alpha$ would have required
a larger set of two-forms and thus be in conflict with 
the field content of the theory (which in turn is determined by supersymmetry).

Let us now repeat this analysis in presence of an $\mathbb{R}^+$-gauging,
i.e.\ for non-vanishing tensor~$\theta_M$.
In this case, the gauge group generators (\ref{defX}) are given by
\bea
X_{MN}{}^K &=&
-\theta^{Li}\,\gamma^{j}{}_{LM}\,(\gamma_{ij})_N{}^K
-\ft1{32}\,\zeta\, (\gamma^{ij})_M{}^L \,(\gamma_{ij})_N{}^K\,\theta_L\,
-\theta_M\,\delta_N^K\;.
\label{X6}
\eea
Using (\ref{relD6}), we obtain for the intertwining tensor $Z$ of (\ref{evalZ})
\bea
Z^{M}{}_{PQ} &=&
-(\gamma_i)_{PQ}\,\Big(\theta^{Mi}+\ft{\zeta}8\,(\gamma^i)^{ML}\,\theta_L\Big)
+(\ft{5\zeta}{16}-1)\,\delta_{(P}^M\,\theta_{Q)}^{\vphantom{K}}
\;.
\eea
Comparing this tensor to (\ref{Z6})
shows that choosing $\zeta=16/5$, $Z$ projects onto the same subspace
(\ref{proj6}) as in absence of the $\mathbb{R}^+$-gauging
\bea
Z^{M}{}_{PQ} &=& (\gamma_i)_{PQ}\,\hat{Z}^{Mi} \;,\qquad
\hat{Z}^{Ki} ~\equiv~ 
-\theta^{Ki}-\ft25\,(\gamma^i)^{KL}\,\theta_L
\;.
\eea
Any other value of $\zeta$ would require a larger set 
of two-forms for consistency of the gauged theory
and thus eventually be inconsistent with supersymmetry.

To summarize, we have found that the presence of a ${\bf 16}_s$ 
component in the embedding tensor~(\ref{dec6}) is possible, if simultaneously
the $\mathbb{R}^+$ trombone symmetry is gauged.
The explicit ansatz for the gauge group generators is
given by (\ref{X6}) with $\zeta=16/5$.
This finishes the discussion of the linear representation constraint
satisfied by the embedding tensor in presence of an 
$\mathbb{R}^+$-gauging.

It remains to evaluate the quadratic constraints (\ref{Q1}), (\ref{Q2})
required for consistency of the gauging.
The constraints~(\ref{Q1}) split into
\bea
(\gamma_i)^{K[M} \,\theta^{N]i}\,\theta_{ K}
&\stackrel!{\equiv}&0\;,
\qquad
\theta^{Ki}\,\theta_K ~\stackrel!{\equiv}~
-\ft25(\gamma^i)^{KL}\,
\theta_{ K} \theta_{ L}
\;,
\label{Q1D6}
\eea
in terms of the irreducible components $\theta^{Ki}$ and $\theta_K$.
These constraints, which are automatically satisfied for $\theta_K=0$
transform under ${\rm SO}(5,5)$
in the ${\bf 120}$ and the ${\bf 10}$ representation, respectively.
Note that the part transforming in the ${\bf 126}_s+{\bf 126}_c$ 
which could in principle be present in (\ref{Q1}) is absent in (\ref{Q1D6}), 
thanks to the particular choice of $\zeta$.
This is crucial for the existence of non-trivial solutions.
Some computation shows that the remaining quadratic 
constraints~(\ref{Q2}) may be cast into the form
\bea
\theta^{Mm}\theta^N_m &\stackrel!{\equiv}& (\gamma^m)^{K(M}\theta^{N)}_m\theta_K
+\ft15\gamma_m^{MN}\theta^{Ki}\theta_K
\;,
\nonumber\\
\theta^{Mi}\theta^{N[k}(\gamma^{l]})_{MN}
&\stackrel!{\equiv}&
\ft1{10}(\gamma^{kl})_M{}^N\theta^{Mi}\theta_N
-\ft35 \theta_M \theta^{M[k}\eta^{l]i}
\;,
\label{Q2D6}
\eea
transforming in the ${\bf 10}\oplus{\bf 126}_c\oplus{\bf 320}$ of ${\rm SO}(5,5)$ and
showing explicitly how the known quadratic constraints of~\cite{Bergshoeff:2007ef}
are modified by the presence of a non-vanishing $\theta_M$.

Every solution $\theta_M$, $\theta^{Mi}$
of the combined set of quadratic constraints (\ref{Q1D6}), (\ref{Q2D6})
will give rise to a consistent gauging of the maximal supergravity in $D=6$.
We have shown that this complete set of constraints transforms as
\bea
{\cal R}_{\rm quad} &=& 
({\bf 10}\oplus{\bf 126}_c\oplus{\bf 320})\oplus
({\bf 10}\oplus{\bf 120}) \;,
\label{qreps6}
\eea
of which the last two representations correspond to (\ref{Q1D6}) and are only relevant 
for a non-vanishing $\theta_M$.
An important non-trivial result in this computation 
(which again hinges on the particular value of 
$\zeta=16/5$ in (\ref{ansatz}) determined above)
is the absence of the ${\bf 1728}$ 
representation in (\ref{Q2D6}) which is a priori possible in (\ref{Q2}). 
As it constitutes the major part of the tensor 
product $\theta^{Mi}\otimes \theta_K$,
a mixed constraint in this representation 
would presumably exclude any solution with both $\theta^{Mi}$ and $\theta_K$ 
non-vanishing.
Instead, we expect a rather rich class of solutions of the quadratic constraints 
(\ref{Q1D6}), (\ref{Q2D6}) with simultaneously non-vanishing~$\theta^{Mi}$ and~$\theta_K$.
We leave the study of such theories to future work.

Let us analyze here in detail the subclass of gaugings with $\theta^{Mi}=0$,
which are thus complementary to the gaugings studied in~\cite{Bergshoeff:2007ef}.
These theories are parametrized by an ${\rm SO}(5,5)$ spinor $\theta_M$  
for which the constraints~(\ref{Q1D6}) reduce to
\bea
(\gamma_i)^{KL}\,\theta_K\theta_L&\equiv&0
\;.
\label{pure6}
\eea
Funny enough, this is precisely the structure of an ${\rm SO}(10)$ {\em pure spinor}
(albeit for a different real form than the usual ${\rm SO}(1,9)$) that shows up 
in a very different context here --- classifying a
particular subsector of possible gaugings in maximal six-dimensional
supergravity.
We can use this to employ the well-known parametrization of the general solution
of this quadratic constraint upon decomposing $\theta_M$ into its 
${\rm GL}(5)$-irreducible parts $(\xi, \xi^m,\xi_{[mn]})$ with $m, n =1, \dots, 5$, 
according to the branching
\bea
{\bf 16}_s &\longrightarrow& 1^{-5} \oplus 5'{}^{+3}\oplus 10^{-1} \;,
\eea
of ${\rm SO}(5,5)$ under ${\rm GL}(5)$. 
In terms of these components, the quadratic constraint 
(alias the pure spinor condition~(\ref{pure6})) 
decomposes into the conditions
\bea
\xi\,\xi^m &=& \epsilon^{mklpq}\,\xi_{kl}\xi_{pq} \;,\qquad
\xi^m\,\xi_{mn} ~=~ 0\;,
\eea
with the totally antisymmetric tensor $\epsilon^{mklpq}$.
On a patch with $\xi\not=0$, these equations are simultaneously solved by setting
\bea
\xi^m= \epsilon^{mklpq}\,\xi_{kl}\xi_{pq}\,/\xi
\;,
\label{solpure6}
\eea
leaving 11 independent real parameters $(\xi,\xi_{mn})$ in the general solution.

We have thus found a particular class of maximal supersymmetric gaugings
defined by $\theta_M=(\xi, \xi^m,\xi_{[mn]})$ with $\xi^m$ given in (\ref{solpure6}).
Moreover, this is the most general gauging with the components $\theta^{Mi}=0$.
As ${\rm GL}(5)$ is the global symmetry group of seven-dimensional
maximal supergravity, it is tempting to speculate that these theories 
have a possible higher-dimensional origin as particular (generalized)
circle compactifications from seven dimensions.
In the following we will see that a very similar pattern shows up for the 
analogous class of gaugings in lower dimensions.


\subsection{$D=5$}


The ungauged theory in $D=5$ dimensions was constructed in~\cite{Cremmer:1980gs},
its general gaugings were given in~\cite{deWit:2004nw}.
The global symmetry group of the ungauged theory
is ${\rm E}_{6(6)}$. Vector and two-form fields of this theory transform in the mutually dual
$\overline{\bf 27}$ and ${\bf 27}$ representation. 
In the ungauged theory, only the vector fields appear in the Lagrangian while the two-forms
are defined as their on-shell duals.
A non-trivial relation among the ${\rm E}_{6(6)}$ generators
that we will exploit in the following, is
\bea
(t^\alpha)_M{}^{K} (t_\alpha)_N{}^{L} &=&
\ft1{18}\,\delta_M^{K}\delta_N^{L} +\ft1{6}\,\delta_M^{L}\delta_N^{K} 
-\ft53\, d_{MNP}\,d^{KLP}
\;,
\label{relD5}
\eea
where $d_{MNK}$ and $d^{MNK}$ are the totally symmetric ${\rm E}_{6(6)}$ invariant tensors,
normalized as $d_{MNK}d^{MNL}=\delta_K^L$, see~\cite{deWit:2004nw}
for further useful relations.

The embedding tensor $\widehat\Theta_M{}^\alpha$ a priori lives
in the tensor product
\bea
{\bf 27} \otimes {\bf 78} &=& 
{\bf 27} \oplus \overline{\bf {351}} \oplus {\bf 1728}
\;,
\label{dec5}
\eea
where ${\bf 78}$ is the adjoint of ${\rm E}_{6(6)}$.
In absence of the $\mathbb{R}^+$-gauging, supersymmetry
restricts the embedding tensor $\Theta_M{}^\alpha$ to the 
$\overline{\bf {351}}$ representation in this decomposition~\cite{deWit:2002vt,deWit:2004nw}.
I.e.\ it can be parametrized in terms of an antisymmetric matrix $Z^{MN}=-Z^{NM}$ as 
\bea
\Theta_{M}{}^{\alpha} &=& 
12\,{Z}^{PQ}\,(t^{\alpha})_{R}{}^{S}\,d^{RKL}d_{MPK}d_{SQL} \;.
\eea
Using (\ref{relD5}), 
we obtain for the full intertwining tensor $Z$ of (\ref{evalZ})
\bea
Z^{M}{}_{PQ} &=& d_{PQL}\,({Z}^{ML}-\ft{5\zeta}3\,d^{MLQ}\theta_{Q})
+(\ft{2\zeta}9-1)\,\delta_{(P}^{\,M}\,\theta_{Q)}^{\vphantom{K}}\;.
\eea
This shows that choosing $\zeta=9/2$, this tensor simplifies to
\bea
Z^{M}{}_{PQ} &=& d_{PQL}\,({Z}^{ML}-\ft{15}2\,d^{MLQ}\theta_{Q})
~\equiv~  d_{PQL}\,\widehat{Z}^{ML}
\;,
\label{Z5}
\eea
and thus projects 
onto a single subrepresentation
\bea
(\overline{\bf 27} \otimes \overline{\bf 27})_{\rm sym} &\longrightarrow& {\bf 27}
\;,
\label{proj5}
\eea
within the full symmetric tensor product, and thus onto the
same subspace as in absence of the $\mathbb{R}^+$-gauging.
This is precisely compatible with the two-forms present in the theory.
Any other value of $\zeta$ would for consistency require a larger set 
of two-forms and thus be incompatible with supersymmetry.

It is interesting to note, that in absence of the $\mathbb{R}^+$-gauging, 
the tensor $\widehat{Z}^{MN}$ of (\ref{Z5}) is totally antisymmetric and 
as such also shows up in the topological coupling of the two-forms in the action
${\cal L}_{B\,dB}=Z^{MN}\,B_M\wedge dB_N$.
The fact that for non-vanishing~$\theta_K$ this tensor is
no longer antisymmetric reflects the fact that the $\mathbb{R}^+$-gaugings 
in general do no longer admit an action.

It remains to evaluate the quadratic constraints (\ref{Q1}), (\ref{Q2})
required for consistency of the gauging.
The quadratic constraints~(\ref{Q1}) split into
\bea
Z^{PQ}\theta_R\,d_{PKM}d_{QLN}d^{KLR} &\stackrel!{\equiv}& 0\;,
\qquad
Z^{MN}\theta_{N}~\stackrel!{\equiv}~ 15\,d^{MKL}\,\theta_K\theta_L
\;,
\label{Q1D5}
\eea
in terms of the irreducible components $Z^{MN}$ and $\theta_K$.
These constraints, which are automatically satisfied for $\theta_M=0$
transform under ${\rm E}_{6(6)}$
in the ${\bf 351}$ and the $\overline{\bf 27}$, respectively.
After some computation, the quadratic constraints~(\ref{Q2}) take the form
\bea
4(t_\alpha)_K{}^L\,Z^{KR}Z^{NS}d_{RSL}
+3(t_\alpha)_K{}^L\,Z^{KN}\theta_L
+3(t_\alpha)_K{}^N\,Z^{KL}\theta_L &\stackrel!{\equiv}& 0
\;.
\label{Q2D5}
\eea
The first term transforms under ${\rm E}_{6(6)}$ in the 
$\overline{\bf 27}+\overline{\bf 1728}$~\cite{deWit:2004nw},
and the form of (\ref{Q2D5}) shows that the additional terms
(upon imposing (\ref{Q1D5})) fall into the same representations.
I.e.\ the total quadratic constraint transforms as
\bea
{\cal R}_{\rm quad} &=& 
(\overline{\bf 27}\oplus\overline{\bf 1728})\oplus
(\overline{\bf 27}\oplus{\bf 351}) \;,
\label{qreps5}
\eea
of which the last two representations correspond to (\ref{Q1D5}) and are only relevant 
in presence of an $\mathbb{R}^+$-gauging.
An important non-trivial result in this constraint analysis 
(which again hinges on the particular value of $\zeta=9/2$ in (\ref{ansatz}))
is the absence of the $\overline{\bf 7371}$ representation which is
a priori possible in (\ref{Q2D5}). As it
constitutes the major part of the tensor product $Z^{MN}\otimes \theta_K$,
its presence among the constraints would presumably exclude any solution 
with both $Z^{MN}$ and $\theta_K$ non-vanishing.

Let us finally discuss the particular gaugings for which $Z^{MN}=0$
and which are thus complementary to those constructed in~\cite{deWit:2004nw}.
In this case, the only non-trivial quadratic constraint on the 
remaining component $\theta_K$ comes from~(\ref{Q1D5}) 
and is given by
\bea
d^{MKL}\,\theta_K\theta_L&\stackrel!{\equiv}& 0
\;.
\label{pure5}
\eea
This condition can be viewed as the ``analogue of a pure spinor'' (\ref{pure6})
for the exceptional group ${\rm E}_{6(6)}$. We can employ a similar 
technique to explicitly solve it. 
To this end, we decompose $\theta_M$ 
into its $({\rm SO}(5,5)\times\mathbb{R}^+)$-irreducible parts 
$(\lambda,\lambda_i,\lambda_\alpha)$ according to the branching
\bea
{\bf 27} &\longrightarrow& 1^{+4} \oplus 10^{-2} \oplus 16_s^{+1} \;.
\eea
The quadratic constraint~(\ref{pure5}) accordingly decomposes into the equations
\bea
\lambda\,\lambda_i &=& (\gamma_i)^{\alpha\beta}\,\lambda_\alpha\lambda_\beta
\;,\qquad
(\gamma^i)^{\alpha\beta}\,\lambda_i\,\lambda_\beta ~=~ 0
\;,\qquad
\lambda_i\, \lambda^i ~=~ 0 
\;,
\eea
with the ${\rm SO}(5,5)$ tensors $ (\gamma_i)^{\alpha\beta}$ introduced in the last
subsection.\footnote{In contrast to the last subsection, we here use indices $\alpha$, $\beta$
for the ${\rm SO}(5,5)$ spinor representation, as capital indices
$M, N$ in this section are reserved for the ${\rm E}_{6(6)}$ fundamental
representation of the vector fields.}
On a patch where $\lambda\not=0$, these equations are simultaneously solved by setting 
\bea
\lambda_i &=& (\gamma_i)^{\alpha\beta}\,\lambda_\alpha\lambda_\beta\,/\lambda\;.
\eea 
This is straightforwardly verified using the well-known identity 
$(\gamma^i)^{(\alpha\beta}(\gamma_i)^{\gamma)\delta}=0$
for ${\rm SO}(5,5)$ gamma-matrices.
In total, this leaves 17 independent real parameters 
$(\lambda, \lambda_\alpha)$ for the general solution of (\ref{pure5}).


\subsection{$D=4$}
\label{subsec:D4}


The ungauged theory in $D=4$ dimensions was constructed in~\cite{Cremmer:1979up},
its general gaugings were given in~\cite{deWit:2007mt}.
The global symmetry group of the ungauged theory
is ${\rm E}_{7(7)}$. Vector and two-form fields of this theory transform in the 
${\bf 56}$ and the adjoint ${\bf 133}$ representations, respectively. 
In the ungauged theory, only 28 electric vector fields appear in the Lagrangian 
while their 28 magnetic duals are defined on-shell. Similarly, the two-forms are
defined on-shell as duals to the scalar fields of the theory.

A non-trivial relation among the ${\rm E}_{7(7)}$ generators
that we will exploit in the following, is
\bea
(t^\alpha)_M{}^{K} (t_\alpha)_N{}^{L} &=&
\ft1{24}\delta_M^{K}\delta_N^{L} + \ft1{12}\delta_M^{L}\delta_N^{K} 
+(t^{\alpha})_{MN}\,(t_{\alpha})^{KL}-\ft1{24}\Omega_{MN}\,\Omega^{KL}
\;,
\label{relD4}
\eea
where the fundamental indices have been raised and
lowered with the symplectic matrix $\Omega_{MN}$
(and we use north-west south-east conventions, i.e. $X^M=\Omega^{MN}X_N$, etc.).

The embedding tensor $\widehat\Theta_M{}^\alpha$ a priori lives
in the tensor product
\bea
{\bf 56} \otimes {\bf 133} &=& 
{\bf 56} \oplus {\bf {912}} \oplus {\bf 6480}
\;.
\label{dec4}
\eea
In absence of the $\mathbb{R}^+$-gauging, supersymmetry
restricts the embedding tensor $\Theta_M{}^\alpha$ to the 
${\bf {912}}$ representation in this decomposition~\cite{deWit:2002vt,deWit:2007mt},
i.e.\ to a tensor satisfying the condition
$\Theta_M{}^\alpha=-2(t_\beta\, t^\alpha)_M{}^N\,\Theta_N{}^\beta$.

Using (\ref{relD4}), 
we obtain for the full intertwining tensor $Z$ of (\ref{evalZ})
\bea
Z^{M}{}_{PQ} &=&
(t_\alpha)_{PQ} \,\Big(
-\ft12 \Theta_{L}{}^\alpha\, \Omega^{ML}
+\zeta\,(t_\alpha)^{ML}\,\theta_L\Big)
+(\ft{\zeta}8-1)\,\delta_{(P}^M\,\theta^{\vphantom{K}}_{Q)}
\;.
\eea
This shows that upon choosing $\zeta=8$, this tensor simplifies to
\bea
Z^{M}{}_{PQ} &=&-\ft12 (t_\alpha)_{PQ} \,\Big(
 \Theta^{M\alpha}
-16\,(t_\alpha)^{ML}\,\theta_L\Big)
~\equiv~  (t_\alpha)_{PQ} \,\,\widehat{Z}^{M\alpha}
\;,
\label{Z4}
\eea
and thus projects 
onto a single subrepresentation
\bea
({\bf 56} \otimes {\bf 56})_{\rm sym} &\longrightarrow& {\bf 133}
\;,
\label{proj4}
\eea
within the full symmetric tensor product, which is the
same subspace as in absence of the $\mathbb{R}^+$-gauging.
This is precisely compatible with the content of two-forms present in the theory.
Any other value of $\zeta$ would for consistency require a larger set 
of two-forms and thus be incompatible with supersymmetry.

It remains to evaluate the quadratic constraints (\ref{Q1}), (\ref{Q2})
required for consistency of the gauging.
The quadratic constraints~(\ref{Q1}) split into
\bea
 (t_{\gamma})_{[M}{}^{ Q}\, \Theta_{N]}{}^\gamma\,\theta_{ Q}
&\stackrel!{\equiv}&0
\;,
\qquad
 \Omega^{PQ}\, \Theta_P{}^\alpha\,\theta_{Q}
~\stackrel!{\equiv}~-16\, (t^\alpha)^{KL} \,\theta_{ K}\, \theta_{ L}
\;,
\label{Q1D4}
\eea
transforming in the ${\bf 1539}$ and the ${\bf 133}$ of 
${\rm E}_{7(7)}$, respectively.
As in the higher dimensions discussed above, the
quadratic constraint (\ref{Q2}) 
in presence of a $\theta_M$ induces a modification of
the known quadratic constraints~\cite{deWit:2007mt} 
which is given by
\bea
\Theta_M{}^\alpha\,\Theta_N{}^\beta\,\Omega^{MN} &\stackrel!{\equiv}&
8 \,\theta_M\,\Theta_N{}^{[\alpha}\,t^{\beta]}{}^{MN}
-4\,f^{\alpha\beta}{}_{\gamma}\,\theta_M\,\Theta_N{}^\gamma\,\Omega^{MN}
\;.
\label{Q2D4}
\eea
Together, we find that the total set of quadratic constraints transforms under
${\rm E}_{7(7)}$ in the representation
\bea
{\cal R}_{\rm quad} &=& 
({\bf 133}\oplus{\bf 8645}) \oplus ({\bf 133}\oplus{\bf 1539})
\;,
\label{qreps4}
\eea
of which the last two representations correspond to (\ref{Q1D4}) and are only relevant 
in presence of a non-vanishing $\theta_M$.

Let us finally discuss the particular gaugings for which $\Theta_M{}^\alpha=0$
and which are thus complementary to those constructed previously in~\cite{deWit:2007mt}.
In this case, the only non-trivial quadratic constraint on the 
remaining component $\theta_K$ comes from~(\ref{Q1D4}) 
and is given by
\bea
(t_\alpha)^{KL}\,\theta_K\theta_L&\stackrel!{\equiv}& 0
\;.
\label{pure4}
\eea
This condition can be viewed as the ``analogue of a pure spinor'' (\ref{pure6})
for the exceptional group ${\rm E}_{7(7)}$. In complete analogy to the analysis for the groups ${\rm SO}(5,5)$ and ${\rm E}_{6(6)}$ above, 
we can find its most general solution by decomposing $\theta_M$ 
into its $({\rm E}_{6(6)}\times\mathbb{R}^+)$-irreducible parts 
$(\eta,\eta_m,\eta^m,\tilde\eta)$
according to the branching
\bea
{\bf 56} &\longrightarrow& 1^{-3} \oplus 27^{-1} \oplus {27}'{}^{+1} \oplus 1^{+3}  \;.
\eea
The quadratic constraint~(\ref{pure4}) accordingly decomposes into the set of equations
\bea
\eta\,\eta^m = d^{mkl}\,\eta_k\eta_l
\;,\qquad
\tilde\eta\,\eta_m = d_{mkl}\,\eta^k\eta^l
\;,\qquad
\eta\tilde\eta -\ft2{15} \eta_m\eta^m = 0= (t_a)_n{}^m\,\eta_m\eta^n
\;,
\eea
with the ${\rm E}_{6(6)}$ tensors $d_{mnk}$
and ${\rm E}_{6(6)}$ generators $(t_a)_n{}^m$
introduced in the last subsection.\footnote{In contrast to the last subsection, 
we here use indices $m$, $n$
for the ${\rm E}_{6(6)}$ fundamental representation, as capital indices
$M, N$ in this section are reserved for the ${\rm E}_{7(7)}$ fundamental
representation of the vector fields.}
On a patch where $\eta\not=0$, these equations are simultaneously solved by setting 
\bea
\eta^m &=& d^{mkl}\,\eta_k\eta_l / \eta \;,
\qquad
\tilde\eta\, ~=~ 
\ft2{15}\, d^{pqr} \eta_p\eta_q\eta_r / \eta^2
\;,
\eea
upon using the identity $d_{p(kl}d_{mn)q}d^{pqr} = \frac2{15}\delta^r_{(k}d^{\vphantom{(}}_{lmn)}$.
This leaves 28 independent parameters $(\eta, \eta_m)$
in the general solution.\footnote{As a byproduct, 
we thus find that an ${\rm E}_{7(7)}$ vector~$\theta_M$ subject to the
quadratic condition~(\ref{pure4}) represents a very compact way
to describe the non-linear conformal realization of this group~\cite{Gunaydin:2000xr}
on a 27-dimensional vector space.}


\subsection{$D=3$}


The ungauged theory in $D=3$ dimensions was constructed in~\cite{Marcus:1983hb},
its general gaugings were given in~\cite{Nicolai:2000sc}.
The global symmetry group of the ungauged theory is ${\rm E}_{8(8)}$. The three-dimensional theory is special in that the ungauged theory does not carry any vector fields which appear in the gauged theory via a Chern-Simons coupling. As they are dual to the scalar fields, they transform in the adjoint  of ${\rm E}_{8(8)}$, which is the ${\bf 248}$-dimensional representation with generators\footnote{In order to facilitate comparison
with previous work in three dimensions~\cite{Nicolai:2000sc,deWit:2008ta}, we use in this section calligraphic indices for the fundamental (=adjoint) representation. Moreover, we use the Cartan-Killing form $\eta_{\cal MN}$ rather than 
the rescaled form $\kappa_{\cal MN}=60\,\eta_{\cal MN}$ defined in footnote~4 
and used in the previous sections
to raise and lower adjoint indices.}
\bea
(t_{\cal M})_{\cal N}{}^{\cal K} &=& -f_{\cal MN}{}^{\cal K}
\;,
\eea
in terms of the structure constants of ${\rm E}_{8(8)}$.
Two forms transform in the ${\bf 1}\oplus{\bf 3875}$ representation of ${\rm E}_{8(8)}$.
Although these forms are non-propagating in three dimensions, their field
content can be inferred from the supersymmetry algebra or from their
on-shell duality to the embedding tensor~\cite{deWit:2008ta,Bergshoeff:2008qd}.

The embedding tensor $\widehat\Theta_{\cal MN}$ a priori lives
in the tensor product
\bea
{\bf 248} \otimes {\bf 248} &=& 
{\bf 1} \oplus {\bf {248}} \oplus {\bf 3875}\oplus {\bf 27000}\oplus {\bf 30380}
\;.
\label{dec3}
\eea
In absence of the $\mathbb{R}^+$-gauging, supersymmetry
restricts the embedding tensor to the reducible
${\bf 1}\oplus{\bf 3875}$ representation in this decomposition~\cite{Nicolai:2000sc}.
Explicitly, this is a symmetric tensor $\Theta_{\cal MN}$ which satisfies
\bea
\Theta_{\cal MN} &=& 
\Big(
({\mathbb P}_{\bf 1})_{\cal MN}{}^{\cal KL} +
({\mathbb P}_{\bf 3875})_{\cal MN}{}^{\cal KL}
\Big)\,\Theta_{\cal KL}
\;,
\eea
with the projectors
\bea 
({\mathbb P}_{\bf 1})_{\cal MN}{}^{\cal KL}     &=& 
\ft{1}{248}\,\eta_{\cal MN} \,\eta^{\cal KL}
\;,\non[.5ex]
({\mathbb P}_{\bf 3875})_{\cal MN}{}^{\cal KL}  &=& 
\ft{1}{7}\,  \delta_{(\cal M}^{\hphantom{(}\cal K} \delta_{\cal N)}^{\cal L} 
 -\ft{1}{14}\, f^{\cal P K}{}_{(\cal M}{} f_{\cal N)P}{}^{\cal L}
 -\ft{1}{56}\, \eta_{\cal MN} \,\eta^{\cal KL}
\;.
\label{projs}
\eea
Using (\ref{ansatz}), 
we obtain for the full intertwining tensor $Z$ of (\ref{evalZ})
\bea
\label{Z3}
Z^{\cal M}{}_{\cal PQ}&=& f^{\cal  L K}{}_{\cal ( P} \Theta_{\cal  Q) L} +
 \left(\zeta f^{\cal  L M}{}_{\cal ( P} f_{\cal  Q) L}{}^{\cal K} 
 - \delta_{\cal  (P}^{\;\,\cal  M} \delta_{\cal  Q)}^{\cal  K}\right) \theta_{\cal  K}
 \;.
\eea
With the explicit form of the projectors~(\ref{projs}) this shows that 
choosing $\zeta=1/2$, the tensor $Z$ projects
onto the subrepresentation
\bea
({\bf 248} \otimes {\bf 248})_{\rm sym} &\longrightarrow& {\bf 1}\oplus{\bf 3875}
\;,
\label{proj3}
\eea
within the full symmetric tensor product, and thus onto the
same subspace as in absence of the $\mathbb{R}^+$-gauging.
This is precisely compatible with the two-forms present in the theory
(which are dual to the embedding tensor)
and thus compatible with supersymmetry 
as we shall explicitly demonstrate in the next section.

It is interesting to note, that in absence of the $\mathbb{R}^+$-gauging, 
the tensor $\widehat{\Theta}_{\cal MN}={\Theta}_{\cal MN}$ is symmetric in its two indices and 
also shows up as a metric of the Chern-Simons term in the action
${\cal L}_{\rm CS}={\Theta}_{\cal MN} A^{\cal M} \wedge dA^{\cal N}$.
The fact that for non-vanishing $\theta_{\cal K}$, this tensor is
no longer symmetric again reflects the fact that the $\mathbb{R}^+$-gaugings 
in general do no longer admit an action.

It remains to evaluate the quadratic constraints (\ref{Q1}), (\ref{Q2})
required for consistency of the gauging which yields
\begin{eqnarray}
\theta_{\cal  M} \theta_{\cal N} - \ft{1}{2} f^{\cal  QP}{}_{\cal  N}f_{\cal  MQ}{}^{\cal  L} \theta_{\cal  P} \theta_{\cal  L} &=&  \Theta_{\cal  LM} f_{\cal  N}{}^{\cal LP} \theta_{\cal  P}
\;,
\label{Q1D3}
\\[1ex]
2 \Theta_{\cal  M L} \Theta_{\cal  T (N} f_{\cal  P)}{}^{\cal L T} - \Theta_{\cal  NP} \theta_{\cal  M} - \Theta_{\cal  L (N} f_{\cal  P)}{}^{\cal L T} f_{\cal  M T}{}^{\cal  Q} \theta_{\cal  Q} &=& 0
\;.
\label{Q2D3}
\end{eqnarray}
In particular, contraction of these equations implies that
\begin{eqnarray}
\theta^{\cal  M} \theta_{\cal  M} &=& 0\;,\qquad
\eta^{\cal MN}\Theta_{\cal M N} \,\theta_{\cal  K} ~=~0~=~
\Theta_{\cal KM}\, \theta^{\cal M}
\;.
\label{QD3trace}
\end{eqnarray}
With some effort one can show that these constraints transform in the
\bea
{\cal R}_{\rm quad} &=&
({\bf 3875} \oplus {\bf 147250}) 
\oplus({\bf 1} \oplus2\cdot{\bf 248} \oplus{\bf 3875} \oplus {\bf 30380})
\;,
\label{qreps3}
\eea
of which the last four representations correspond to (\ref{Q1D3}) 
and (\ref{QD3trace}) and are only relevant for a non-vanishing $\theta_{\cal M}$.
Notably, the ${\bf 779247}$ representation which is not excluded
by group theory arguments and could in principle show up
among these constraints is explicitly absent.
We shall come back to (a proof of) this fact in the next section.

The second equation in (\ref{QD3trace}) implies that
the singlet ${\bf 1}$ and the vector ${\bf 248}$ component
of the embedding tensor cannot be switched on simultaneously.
Absence of the vector $\theta_{\cal M}$ corresponds to the theories 
without gauging of the ${\mathbb R}^+$ scaling symmetry.
As these theories have been discussed in detail in~\cite{Nicolai:2000sc},
we shall in the following assume a non-vanishing vector $\theta_{\cal M}$
and thus a vanishing singlet component $\eta^{\cal MN}\Theta_{\cal M N} $
of the three-dimensional embedding tensor.

Let us finally discuss the particular gaugings for which $\Theta_{\cal M N}=0$
and which are thus complementary to those constructed in~\cite{Nicolai:2000sc}.
In this case, the only non-trivial quadratic constraint on the 
remaining component $\theta_K$ comes from~(\ref{Q1D3}) 
and is given by
\bea
\Big(
({\mathbb P}_{\bf 1})_{\cal MN}{}^{\cal KL} +
({\mathbb P}_{\bf 3875})_{\cal MN}{}^{\cal KL}
\Big)\;\theta_{\cal K}\,\theta_{\cal L} &\stackrel!{\equiv}& 0
\;.
\label{pure3}
\eea
As in higher dimensions,
this condition can thus be viewed as the ``analogue of a pure spinor'' (\ref{pure6})
for the exceptional group ${\rm E}_{8(8)}$. 
We can use the same technology in order to find its general solution.
As the calculation is somewhat more involved that for the higher-dimensional
cases, we defer the details to appendix~\ref{app:constraint} and just present
the solution here. Decomposing $\theta_{\cal M}$ under 
${\rm E}_{7(7)}\times \mathbb{R}^+$ into components
$(\tilde\eta, \tilde\eta_m, \xi, \xi_{\alpha}, \eta_m, \eta)$ according to the decomposition\footnote{Here, we use indices $m$ and $\alpha$ 
for the fundamental 56 and the adjoint 133 of ${\rm E}_{7(7)}$, respectively.}
\bea
\mathbf{248} &\longrightarrow& {1}^{+2} \oplus{56}^{+1} \oplus{1}^{0} \oplus {133}^0  \oplus {56}^{-1}\oplus {1}^{-2}
\;,
\label{break248R}
\eea
the general solution of (\ref{pure3}) can be expressed in terms of the 
58 parameters $\eta$, $\eta_m$, $\xi$ as
\bea
\xi_{\alpha} &=& - \frac{6}{\eta} \, (t_{\alpha})^{mn} \, \eta_m \, \eta_n
\;,
\nonumber\\[.5ex]
\tilde\eta_m &=& \frac{\xi}{\eta} \, \eta_m  - \frac{24}{\eta^2} \, 
(t^{\alpha})_m{}^{n} \, (t_{\alpha})^{pq} \, \eta_n \, \eta_p \, \eta_q \;,
\nonumber\\[.5ex]
\tilde\eta &=& \frac{\xi^2}{\eta} - \frac{2}{\eta} \, \xi^{\alpha} \, \xi_{\alpha}\;.
\label{sol3}
\eea
We note that the second term in $\tilde\eta$ is related 
to the quartic invariant 
$(t_{\alpha})^{kl}(t^{\alpha})^{mn} \, \eta_k \eta_l \eta_m \eta_n$
of ${\rm E}_{7(7)}$.
Like for ${\rm E}_{7(7)}$ above, 
the explicit solution (\ref{sol3}) in terms of 58 parameters 
shows that a vector subject to the bilinear condition (\ref{pure3})
represents a very compact way
to describe the non-linear conformal realization of ${\rm E}_{8(8)}$
given in~\cite{Gunaydin:2000xr}.


\subsection{Summary}


We have in this section explicitly constructed the gauge group generators of the 
gaugings of maximal supergravity that involve also a gauging of the 
on-shell scaling symmetry~(\ref{scaling}).
In dimensions $3\le D\le 6$, these generators are given by (\ref{ansatz})
with the respective values of $\zeta$ computed above in the various subsections.
The possibility of a local scaling symmetry gives rise to another set of 
parameters  $\theta_M$
within the embedding tensor that transform in the dual vector representation.
We have worked out for all cases the quadratic constraints on the embedding
tensor required for consistency.
In particular, for those gaugings that are exclusively triggered by the 
new parameters $\theta_M$, we have furthermore given the explicit solution 
of these consistency constraints in all cases.

While so far we have only derived the necessary algebraic consistency constraints,
it remains to show that every solution to these constraints (e.g.\ to
equations~(\ref{Q1D4}), (\ref{Q2D4}) in $D=4$ dimensions) indeed gives rise to
a consistent theory. In particular, it remains to determine the deformed
field equations --- as the theory no longer admits an action, the analysis
must be performed on the level of the equations of motion. 
This will be the subject of the next section. We pick the example of the
maximal $D=3$ supergravity, for which the algebraic structure is the 
most involved one, and show how the equations of motion must be modified
under gauging in order to remain supersymmetric.


\section{Supersymmetry}

\label{sec:supersymmetry}


In this section we will take as an example the maximal
three-dimensional theory and work out the full set of 
the deformed equations of motion. In particular,
this will show that the quadratic constraints~(\ref{Q1D3}), (\ref{Q2D3})
are sufficient for consistency of the theory, in other words,
that every solution to these equations defines a 
consistent and maximally supersymmetric gauging in three dimensions.
Upon dimensional reduction
the algebraic structures which connect gauging and supersymmetry
are embedded into the increasing symmetry algebras.
The results of this section thus give some strong evidence that also the
algebraic constraints we have derived in 
sections~\ref{subsec:D6} -- \ref{subsec:D4}
for the higher dimensions are
sufficient for compatibility with supersymmetry.

As we have repeatedly mentioned, the resulting theory does not
admit an action. The analysis must therefore be performed on the level of 
the equations of motion. After reviewing the three-dimensional
theory we analyze the deformed supersymmetry algebra and 
in section~\ref{subsec:eqm} we derive the full set
of the deformed equations of motion (to lowest order in the fermions).


\subsection{The three-dimensional theory}


We recall some basic notations of the maximal three-dimensional 
supergravity and its gaugings, 
see~\cite{Marcus:1983hb,Nicolai:2000sc}
for details. Also we have collected in appendix~\ref{app:e8} our conventions
for the exceptional group ${\rm E}_{8(8)}$.

The scalar fields in three dimensions are described by an ${\rm E}_{8(8)}$-valued
matrix ${\cal V}^{\cal M}{}_{\underline{\cal M}}$, with the two indices labeling
the  248-dimensional adjoint representation and indicating the transformation properties
\bea
\delta {\cal V} &=& \Lambda {\cal V} - {\cal V} h(x)
\;, \qquad
\Lambda\in\mathfrak{e}_{8(8)}\;,\quad 
h(x)\in \mathfrak{so}(16)
\;,
\eea
under global ${\rm E}_{8(8)}$ and local ${\rm SO}(16)$, respectively.
In particular, it is customary to split the group matrix according to
${\cal V}^{\cal M}{}_{\underline{\cal M}} = \{{\cal V}^{\cal M}{}_{IJ},
{\cal V}^{\cal M}{}_{A}\}$, according to the decomposition
of $\mathfrak{e}_{8(8)}$ into its compact subalgebra 
$\mathfrak{so}(16)=\langle X^{[IJ]}\rangle$
and 128 noncompact generators $\{ Y^A \}$.
Here
$I,J,\dots$ and $A, B,\dots$, respectively, label the ${\bf 16}$ and
${\bf 128}_s$ representations of SO(16). Eventually we will also need
indices $\dot A,\dot B,\dots$ labelling the conjugate spinor
representation ${\bf 128}_c$. Naturally we will also encounter
$\mathrm{SO}(16)$ gamma matrices $\Gamma^I_{A\dot A}$ in what
follows. We will freely raise and lower $\mathrm{SO}(16)$ indices.

In this basis, the Cartan-Killing form $\eta_{\cal MN}$ of ${\rm E}_{8(8)}$ takes 
the form
\bea
\eta_{\cal MN}\,
{\cal V}^{\cal M}{}_{IJ}
{\cal V}^{\cal N}{}_{KL}
=
-2\delta^{IJ}_{KL}
\;,
\qquad
\eta_{\cal MN}\,
{\cal V}^{\cal M}{}_{A}
{\cal V}^{\cal N}{}_{B}
=
\delta_{AB}
\;,
\label{CK}
\eea
and the ${\rm E}_{8(8)}$ structure constants $f^{\cal MNK}$ 
can be expressed as
\bea
f^{\cal MNK}
&=&
{\cal V}^{\cal M}{}_{\underline{\cal M}}
{\cal V}^{\cal N}{}_{\underline{\cal N}}
{\cal V}^{\cal K}{}_{\underline{\cal K}}\,
f^{\underline{\cal MNK}}
\nonumber\\[1ex]
&=&
-
{\cal V}^{\cal M}{}_{IJ}
{\cal V}^{\cal N}{}_{KL}
{\cal V}^{\cal K}{}_{MN}\,\Big(
\delta^{IK}\delta^{LM}\delta^{NJ}
\Big)
-
\ft34
{\cal V}^{[\cal M}{}_{IJ}
{\cal V}^{\cal N}{}_{A}
{\cal V}^{\cal K]}{}_{B}
\Big(
\Gamma^{IJ}_{AB}
\Big)
\;.
\label{fff}
\eea
The inverse matrix ${\cal V}^{\underline{\cal M}}{}_{\cal M}$ is defined 
by\footnote{Note that these conventions gives rise to the relations
${\cal V}_{\cal M}{}^{IJ}\equiv \eta_{\cal MN}{\cal V}^{{\cal N} IJ}
=-{\cal V}^{IJ}{}_{\cal M}$ and
${\cal V}_{\cal M}{}^{A}\equiv \eta_{\cal MN}{\cal V}^{{\cal N} A}
={\cal V}^{A}{}_{\cal M}$, cf.~appendix~\ref{app:e8}.
}
\bea
{\cal V}^{KL}{}_{\cal M}\,{\cal V}^{\cal M}{}_{IJ}~=~2\delta_{KL}^{IJ}\;,
\qquad
{\cal V}^{A}{}_{\cal M}\,{\cal V}^{\cal M}{}_{B}~=~\delta_{B}^{A}
\;.
\eea

The standard gaugings are defined in terms of the embedding tensor 
$\Theta_{\cal MN}$. The fermionic mass terms of the theory as well as 
the scalar potential can be expressed in terms of the $T$-tensor
\bea
T_{\underline{\cal M}|\underline{\cal N}}
&\equiv &
\Theta_{\cal MN}
\,{\cal V}^{{\cal M}}{}_{\underline{\cal M}}{}{\cal V}^{{\cal N}}{}_{\underline{\cal N}}
\;,
\label{Told}
\eea
obtained by dressing the embedding tensor with the scalar matrix ${\cal V}^{{\cal M}}{}_{\underline{\cal M}}$. Similarly, the crucial object in the description of the gaugings with local scaling symmetry will be the dressed new component $\theta_{\cal M}$:
\bea
T_{\underline{\cal M}}
&\equiv &
\theta_{\cal M}\,
{\cal V}^{{\cal M}}{}_{\underline{\cal M}}
\;.
\label{Tnew}
\eea
As $\Theta_{\cal MN}$ is restricted to live in the ${\bf 1}\oplus{\bf 3875}$ representation
of ${\rm E}_{8(8)}$, the same applies to the $T$-tensor. It can hence be expressed 
as
\bea
\Theta_{\cal MN}
&=&
{\cal V}^{\underline{\cal M}}{}_{\cal M}{}{\cal V}{}^{\underline{\cal N}}{}_{\cal N}
\,T_{\underline{\cal M}|\underline{\cal N}}
\nonumber\\[1.5ex]
&=&
\ft14\,{\cal V}_{\cal M}{}^{IJ}
{\cal V}_{\cal N}{}^{KL}
\,\Big( 
\delta_{\vphantom{[}}^{I[K} A_1^{L]J}- \delta_{\vphantom{[}}^{J[K} A_1^{L]I}
+\ft1{64} \Gamma^{IJKL}_{\dot{A}\dot{B}}\,A_3^{\dot{A}\dot{B}} \Big)
\non
&&{}-
{\cal V}_{(\cal M}{}^{IJ}
{\cal V}_{\cal N)}{}^{A}
\,\Big( 
 \Gamma^{[I}_{A\dot{A}}\,A_2^{J]\dot{A}}
\Big)
~+~
{\cal V}_{\cal M}{}^{A}
{\cal V}_{\cal N}{}^{B}
\,\Big(  
\ft1{16}\,\Gamma^I_{A\dot{A}}\Gamma^I_{B\dot{B}}\,A_3^{\dot{A}\dot{B}} 
\Big)
\;,
\label{A123}
\eea
in terms of three tensors $A_1$, $A_2$ and $A_3$ transforming in the
${\bf 135}$, ${\bf 1920}_c$ and ${\bf 1820}$
of ${\rm SO}(16)$, respectively, i.e.\ satisfying
\bea
A_1^{IJ}~=~A_1^{JI}\,\qquad
\Gamma^I_{A\dot{A}}\,A_2^{I\dot{A}}~=~0\;,\qquad
A_3^{\dot{A}\dot{B}}~=~
\ft1{3072}\,
\Gamma^{IJKL}_{\dot{A}\dot{B}}\Gamma^{IJKL}_{\dot{C}\dot{D}}\,A_3^{\dot{C}\dot{D}}
\;.
\eea
In the standard gauged theory (in absence of the $\mathbb{R}^+$-gauging), 
these terms describe the various fermionic
mass term in the Lagrangian while the scalar potential is given by
\bea
W(\phi) &=& \ft{1}{4} \, g^2 \, \Big(
A_2^{I \dot A} A_2^{I \dot A}-2 A_1^{IJ} A_1^{IJ} 
 \Big)
\;.
\label{potentialold}
\eea

Similarly, we now introduce tensors $B_{IJ}$ and $B_{A}$ in order to
parametrize the different ${\rm SO}(16)$ components of the new 
part~(\ref{Tnew}) of the $T$-tensor
\bea
\theta_{\cal M} &=& {\cal V}_{\cal M}{}^{\underline{\cal M}}\,T_{\underline{\cal M}}~=~
\ft12\, {\cal V}_{\cal M}{}^{IJ}\,B_{IJ} +{\cal V}_{\cal M}{}^{A}\,B_{A}
\;.
\label{AB}
\eea

With the currents
\bea
{\cal V}^{-1}(\partial_\mu-g A^{\cal M}_\mu \,\widehat{\Theta}_{\cal MN}\, t^{\cal N})\,{\cal V} &\equiv&
\ft12 {\cal Q}_\mu^{IJ}X^{IJ}+{\cal P}_\mu^A Y^A
\;,
\label{PQ}
\eea
we find among the various components of the $T$-tensor the
differential relations
\bea
{\cal D}_\mu A_1^{IJ} &=&
\Gamma^{(I}_{A\dot A}\,A_2^{J)\dot A}\,{\cal P}^A_\mu 
\;, 
\non
{\cal D}_\mu A_2^{I\dot A} &=&
\ft12\,\left(
\Gamma^M_{A\dot A}\,A_1^{IM} + \Gamma^I_{A\dot B}\,A_3^{\dot A\dot B} 
-\ft1{16}\, (\Gamma^I
\Gamma^J)_{\dot{A}\dot{B}}\Gamma^J_{A\dot{C}}\,A_3^{\dot{C}\dot{B}} \right)\,{\cal P}^A_\mu 
\;,
\non[.5ex]
{\cal D}_\mu B_{IJ} &=& \ft12 \Gamma^{IJ}_{AB}\, B_A\,{\cal P}^B_\mu\;,\qquad
{\cal D}_\mu B_A ~=~ \ft14  \Gamma^{IJ}_{AB}\, B_{IJ}\,{\cal P}^B_\mu\;,
\label{ABderivatives}
\eea
where ${\cal D}_\mu$ denotes the full ${\rm SO}(16)$-covariant derivative.


\subsection{Implications of the quadratic constraint}
\label{subsec:quadratic}


In this section we will compute and collect a number of relations
that can be derived from the quadratic constraints~(\ref{Q1D3}), (\ref{Q2D3})
on the embedding tensor.
The section is largely technical and since the algebraic calculations become quite
involved we have made repeated use of the computer algebra system 
Cadabra~\cite{Peeters:2006kp} to organize 
and simplify the computation.

We have seen that the gauging of the theory is described in terms
of the embedding tensor, which is parametrized by components 
$\Theta_{\cal MN}$, $\theta_{\cal M}$, subject to the relations~(\ref{Q1D3}), (\ref{Q2D3}).
The equations of motion of the theory on the other hand feature the dressed
version of the embedding tensor defined in (\ref{Told}) and (\ref{Tnew}).
In order to appreciate the consequences of the quadratic constraint, we will
thus have to translate equations~(\ref{Q1D3}), (\ref{Q2D3}) into relations between the scalar dependent tensors $A_{1,2,3}$ and $B$ from (\ref{A123}), (\ref{AB}).

Let us start from the simplest set of constraints (\ref{QD3trace}). 
Its second equation translates into
\bea
\eta^{\cal MN}\Theta_{\cal MN}\,B_A &=& 0 ~=~ 
\eta^{\cal MN}\Theta_{\cal MN}\,B_{IJ}
\;,
\eea
and as mentioned above, it is automatically solved if 
$\Theta_{\cal MN}$ transforms in the ${\bf 3875}$ and has no singlet component.
Plugging the explicit expansions (\ref{A123}), (\ref{AB}) 
into the remaining equations of (\ref{QD3trace}) 
gives rise to the relations
\bea
B_{IJ} B_{IJ} - 2\, B_A B_A &=& 0
\;,\non[.5ex]
A_1^{K[I}\,B_{\vphantom{1}}^{J]K}+\Gamma^{[I}_{A\dot{A}}\,A_2^{J]\dot{A}}\,B_A
-\ft1{128}\,\Gamma^{IJKL}_{\dot{A}\dot{B}}\,A_3^{\dot{A}\dot{B}}\,B_{KL}
&=& 0
\;,
\nonumber\\
8\,\Gamma^{I}_{A\dot{A}}\,A_2^{J\dot{A}}\,B_{IJ}
-\Gamma^I_{A\dot{A}}\Gamma^I_{B\dot{B}}\,A_3^{\dot{A}\dot{B}}\,B_B &=& 0
\;.
\label{3-3}
\eea
We note that these constraints transform in the ${\bf 1}$, ${\bf 120}$, and ${\bf 128}_s$ 
of ${\rm SO}(16)$, respectively.
On the other hand,
from evaluating (\ref{Q1D3}) for ${\cal M}=[IK]$ and ${\cal N}=[JK]$ we obtain
after subsequent symmetrisation and antisymmetrisation in $I, J$,
two equations in the ${\bf 135}$, and ${\bf 120}$
\bea
8 B_{IK}B_{JK}+8A_1^{K(I}B_{\vphantom{2}}^{J)K}
-4 \Gamma^{(I}_{A\dot{A}}\,A_2^{J)\dot{A}}\,B_A - \delta_{IJ}\,B_AB_A &=& 0
\;,
\nonumber\\
-6A_1^{K[I}B_{\vphantom{2}}^{J]K}
-3\Gamma^{[I}_{A\dot{A}}\,A_2^{J]\dot{A}}\,B_A
+\ft1{64}\,\Gamma^{IJKL}_{\dot{A}\dot{B}}\,A_3^{\dot{A}\dot{B}}\,B_{KL}
&=& 0
\;,
\label{3-32}
\eea
respectively. Instead, choosing in (\ref{Q1D3}) ${\cal M}=[IJ]$ and ${\cal N}=A$ and
contracting the equation with $\Gamma^J_{A\dot{A}}$ leads to
\bea
0&=\!& 7\Gamma^J_{A\dot{A}}\,B_{IJ} B_A
-\ft7{16}\Gamma^I_{A\dot{A}}\,\Gamma^{JK}_{AB}\,B_{JK}B_{B}
+\ft72\Gamma^J_{A\dot{A}}\,A_1^{IJ} B_A-A_2^{J\dot{A}}\,B_{IJ}
+\ft32\Gamma^I_{A\dot{B}}\,A_3^{\dot{A}\dot{B}}\,B_A
\nonumber\\
&&{}-\ft12\,(\Gamma^I\Gamma^K)_{\dot{A}\dot{B}}\,A_2^{J\dot{B}}B_{JK}
-\ft54\,\Gamma^{JK}_{\dot{A}\dot{B}}\,A_2^{I\dot{B}}B_{JK}
-\ft3{16}(\Gamma^I \Gamma^J)_{\dot A\dot{B}}\Gamma^J_{B\dot{C}}\,
A_3^{\dot{B}\dot{C}}\,B_B
\;.
\label{18-1}
\eea
Upon interchanging ${\cal M}$ and ${\cal N}$ in (\ref{Q1D3}), the same contraction
yields
\bea
0&=& 7\Gamma^J_{A\dot{A}}\,B_{IJ} B_A
-\ft7{16}\Gamma^I_{A\dot{A}}\,\Gamma^{JK}_{AB}\,B_{JK}B_{B}
-6A_2^{J\dot{A}}\,B_{IJ}
+2\Gamma^I_{A\dot{B}}\,A_3^{\dot{A}\dot{B}}\,B_A
\nonumber\\
&&{}+\ft12\,(\Gamma^I\Gamma^K)_{\dot{A}\dot{B}}\,A_2^{J\dot{B}}B_{JK}
-\ft12\,\Gamma^{JK}_{\dot{A}\dot{B}}\,A_2^{I\dot{B}}B_{JK}
-\ft1{32}(\Gamma^I\Gamma^J)_{\dot A\dot{B}}\Gamma^J_{B\dot{C}}\,
A_3^{\dot{B}\dot{C}}\,B_B
\;.
\qquad
\label{18-2}
\eea
Under ${\rm SO}(16)$ the two equations (\ref{18-1}) and (\ref{18-2})
transform in the ${\bf 128}_s\oplus{\bf 1920}_c$ and it is 
straightforward to verify that the two parts in the ${\bf 128}_s$ 
(obtained by further contraction with $\Gamma^I_{B\dot{A}}$)
are proportional to the last equation of (\ref{3-3}).

Finally, we evaluate part of the quadratic constraint (\ref{Q2D3}). 
Choosing ${\cal M}=[JK]$, ${\cal N}=[IM]$, ${\cal P}=[KM]$
and symmetrizing in $(IJ)$ leads to the relation
\bea
0 &=& 
A_1^{IK}A_1^{JK}-\ft12A_2^{I\dot A}A_2^{J\dot A} 
+A_1^{K(I}B_{\vphantom{2}}^{J)K}-\ft14\Gamma^{(I}_{A\dot A}A_2^{J)\dot A} B_A 
\non
&&{}
-\ft1{16}\,\delta^{IJ}\,(A_1^{KL}A_1^{KL}-\ft12A_2^{K\dot A}A_2^{K\dot A} )
\;,
\label{4-1}
\eea
in the ${\bf 135}$ of ${\rm SO}(16)$.
Choosing in (\ref{Q2D3})
${\cal M}=A$, ${\cal N}=[IM]$, ${\cal P}=[KM]$ and contracting with $\Gamma^K_{A\dot{A}}$
we obtain
\bea
\ft1{64}\,\Gamma^{IJKL}_{\dot{C}\dot{D}}\Gamma^{KL}_{\dot{A}\dot{B}}\,
A_2^{J\dot{B}}{}A_3^{\dot{C}\dot{D}}
&=&
-32A_1^{IJ}A_2^{J\dot A}
+2(\Gamma^I\Gamma^J)_{\dot A\dot B} \,A_1^{JK}A_2^{J\dot B} 
+10 A_2^{I\dot B} A_3^{\dot A\dot B}
\non[1ex]
&&{}
- (\Gamma^I\Gamma^J)_{\dot A\dot B} \,A_2^{J\dot C}A_3^{\dot B\dot C}
+20\Gamma^J_{A\dot A}\,A_1^{IJ}B_A
+2 A_2^{J\dot A}B_{IJ}
\non[1ex]
&&{}
-\ft72 \Gamma^{JK}_{\dot A\dot B}\,A_2^{I\dot B}\,B_{JK}
+(\Gamma^I\Gamma^K)_{\dot A\dot B}\,A_2^{J\dot B}B_{JK} 
\;.
\eea
Choosing instead 
${\cal M}=[IM]$, ${\cal N}=[KM]$, ${\cal P}=A$ before contracting with $\Gamma^K_{A\dot{A}}$,
we obtain
\bea
\ft1{64}\,\Gamma^{IJKL}_{\dot{C}\dot{D}}\Gamma^{KL}_{\dot{A}\dot{B}}\,
A_2^{J\dot{B}}{}A_3^{\dot{C}\dot{D}}
&=&
64A_1^{IJ}A_2^{J\dot A}
-4(\Gamma^I\Gamma^J)_{\dot A\dot B} \,A_1^{JK}A_2^{J\dot B} 
-22 A_2^{I\dot B} A_3^{\dot A\dot B}
\non[1ex]
&&{}
+ (\Gamma^I\Gamma^J)_{\dot A\dot B} \,A_2^{J\dot C}A_3^{\dot B\dot C}
+\Gamma^J_{A\dot A}\,A_1^{IJ}B_A
+64 A_2^{J\dot A}B_{IJ}
\non[1ex]
&&{}
-2(\Gamma^I\Gamma^K)_{\dot A\dot B}\,A_2^{J\dot B}B_{JK} 
-11\Gamma^I_{A\dot B}\,A_3^{\dot A\dot B}\,B_A
\non[1ex]
&&{}
-\ft1{16}(\Gamma^I\Gamma^J)_{\dot A\dot B}
\Gamma^J_{A\dot C}\,A_3^{\dot B\dot C}\,B_A
\;.
\label{4-3}
\eea
Again these two equations transform in the ${\bf 128}_s\oplus{\bf 1920}_c$
and one verifies that both ${\bf 128}_s$ parts reduce to the last equation of~(\ref{3-3}).
We note that in absence of the vector $\theta_{\cal M}$,
(i.e.\ for $B_{IJ}=0=B_A$)
all these equations consistently reduce to equations 
(4.17) and (4.19) of~\cite{Nicolai:2000sc}.
Together, we have thus shown that the lowest ${\rm SO}(16)$ representations
appearing in the quadratic constraint are given by
\bea
{\cal R}_{\rm quad} &=&
{\bf 1}\oplus3\cdot{\bf 120}\oplus2\cdot{\bf 128}_s
\oplus3\cdot{\bf 135}\oplus4\cdot{\bf 1920}_c\oplus\dots
\;,
\eea
in agreement with the corresponding decomposition 
of (\ref{qreps3}). In particular, the fact that within all
the above equations there are only two independent constraints 
in the ${\bf 128}_s$ finally proves that there is no ${\rm E}_{8(8)}$
representation ${\bf 779247}$ in the quadratic constraint~(\ref{qreps3}).
Its presence would have excluded all solutions to the quadratic constraint
with both $\Theta_{\cal MN}$ and $\theta_{\cal M}$ non-vanishing.

In order to study supersymmetry of the equations of motion
in the next section, we will need the following particular linear
combinations of the above constraints in the 
the ${\bf 128}_s\oplus{\bf 1920}_c$
representation
\bea
0&=&
3\, A_1^{I J} A_2^{J \dot A} 
- A_2^{I \dot B} A_3^{\dot A \dot B} 
+ 3\, A_2^{J \dot A} B_{I J} 
- \ft{1}{4}\, A_2^{I \dot B} B_{J K} \Gamma^{JK}_{\dot A \dot B} 
+ \ft{3}{4}\, A_1^{I J} B_{A} \Gamma^J_{A \dot A} 
\non[1ex]
&&{}
- \ft{1}{4}\, A_3^{\dot A \dot B} B_{A} \Gamma^I_{A \dot B} 
+ B_{A} B_{I J} \Gamma^J_{A \dot A} 
- \ft{3}{16}\, A_1^{J K} A_2^{K \dot B} (\Gamma^I\Gamma^{J})_{\dot A \dot B} 
+ \ft{1}{16}\, A_2^{J \dot C} A_3^{\dot B \dot C} (\Gamma^I\Gamma^J)_{\dot A \dot B} 
\non[1ex]
&&{}
- \ft{1}{4}\, A_2^{K \dot B} B_{J K} (\Gamma^I\Gamma^J)_{\dot A \dot B} 
+ \ft{1}{64}\, A_3^{\dot B \dot C} B_{A} (\Gamma^{I}\Gamma^{J})_{\dot A \dot B} \Gamma^J_{A \dot C} 
- \ft{1}{16}\, B_{A} B_{J K} (\Gamma^I \Gamma^{JK})_{\dot A A} 
\;.
\non[1ex]
\label{final2}
\eea
Again, in absence of $B_{IJ}$ and $B_A$
this equation consistently reduces to the constraint
derived in~\cite{Nicolai:2000sc}, section 4.4.


\subsection{Supersymmetry algebra}


We will now study the effect of the gauging on the 
three-dimensional supersymmetry algebra. This will allow us to derive the
deformed supersymmetry transformation rules which we will subsequently
use to determine the full set of deformed field equations.
For the standard gaugings (in absence of the vector $\theta_{\cal M}$),
the supersymmetry algebra in three dimensions has
recently been computed for all $p$-forms~\cite{deWit:2008ta}.

For the bosonic fields $e_\mu{}^\alpha$, ${\cal V}$ and $A_\mu^{\cal M}$,
the supersymmetry transformation rules are given by
\begin{eqnarray}
  \delta {e_\mu}^\alpha &=& 
  \mathrm{i} \bar{\epsilon}^I \gamma^\alpha \psi_\mu{}^I 
  \;,\qquad
  {\cal V}^{-1} \delta {\cal V} ~=~ 
  \Gamma^I_{A\dot A} \,\bar{\chi}^{\dot A} \epsilon^I \,Y^A
  \;,\nonumber\\[.5ex]
  \delta A_\mu{}^{\cal M} &=&
  2 \,{\cal V}^{\cal M}_{\;\;\;\;IJ}\,{\bar{\epsilon}}^I\psi_\mu{}^J 
  -\mathrm{i} \Gamma^I{}_{A\dot A}\,{\cal V}^{\cal
    M}_{\;\;\;\;A}\;{\bar{\epsilon}}^I\gamma_{\mu}\chi^{\dot A}  
  \;,
    \label{susyB}
\end{eqnarray}
and do not change upon gauging.\footnote{Our 
space-time conventions are a signature $(+--)$
for the three-dimensional metric 
$g_{\mu\nu}$, and
$e\gamma^{\mu\nu\rho}=-\mathrm{i}\varepsilon^{\mu\nu\rho}$
for the ${\rm SO}(1,2)$ $\gamma$-matrices.}
The fermionic fields appearing
in these transformations are 16 gravitinos $\psi_\mu{}^I$ and 128 spin-1/2
fermions $\chi^{\dot A}$ transforming under ${\rm SO}(16)$. In the
presence of a gauging their supersymmetry variations are given 
(up to higher order fermionic contributions)
by
\begin{eqnarray}
{\delta_{\epsilon}} \psi_{\mu}^I &=& 
{\cal D}_{\mu} \epsilon^{I} + 
i g \left(A_1 ^{IJ}+\beta_1\,B_{IJ}\right) \gamma_{\mu} \epsilon^J  
\;,
\nonumber\\
{\delta_{\epsilon}} \chi^{\dot A} &=& \frac{i}{2} \gamma^{\mu} \epsilon^I \Gamma^{I}_{A {\dot A}} {\cal P}_{\mu}^{A} + g 
\left( A_2 ^{I {\dot A}}+\beta_2\, \Gamma^I _{A {\dot A}} B_A \right) \epsilon^I
\;,
\label{susyF}
\end{eqnarray}
with the tensors $A_{1}$, $A_{2}$ and $B$ from (\ref{A123}), (\ref{AB}) above
and some constants $\beta_{1,2}$.
The covariant derivative is explicitly given by
\bea
{\cal D}_{\mu} \epsilon^{I}  &=&
(\partial_\mu  + \ft14 \widehat\omega_{\mu}{}^{ab}\,\gamma_{ab}
-\ft12 {\cal A}_\mu )\,\epsilon^I 
+  {\cal Q}_\mu^{IJ}\epsilon^J
\;.
\label{covexplicit}
\eea
The effect of a gauging with non-vanishing vector $\theta_{\cal M}$ in 
these transformations is furthermore reflected by the terms in $B_{IJ}$ and $B_A$ 
which are entirely determined by their
index structure up to the global factors~$\beta_{1,2}$.
The latter are fixed by demanding closure of the supersymmetry algebra
into diffeomorphisms, Lorentz transformations, ${\rm SO}(16)$ transformations
and gauge transformations:
\begin{eqnarray}
  {}[\delta_{\epsilon_{1}},\delta_{\epsilon_{2}}] &=&
  \delta_\xi +\delta_\omega+\delta_h +  \delta_{\Lambda}
   \;. 
     \label{susyalgebra}
\end{eqnarray}
Setting $\beta_1=-1$, $\beta_2=\ft14$ , one can verify that the supersymmetry 
transformations (\ref{susyB}), (\ref{susyF}) close on the vielbein $e_\mu{}^\alpha$
and on the scalar fields ${\cal V}$ into the
algebra (\ref{susyalgebra}) with diffeomorphism and gauge
parameter given by
\begin{eqnarray}
  \label{susypars}
  \xi^{\mu}&=&-\mathrm{i}\,
  \bar\epsilon_{[1}{}^I\gamma^{\mu}\epsilon_{2]}{}^{I} \;,
  \nonumber\\[1ex]
  \Lambda^{\cal M}&=&
  -\xi^{\rho}A_{\rho}^{\cal M}
  -2\,{\cal V}^{\cal M}{}_{IJ}\,
  \bar\epsilon_{[1}{}^{I}\epsilon_{2]}{}^J \;.
\end{eqnarray}
On the vector fields, the commutator of two supersymmetry
transformations yields
(again up to higher order fermionic terms)
\bea
{}
[ \delta_{\epsilon_{1}},  \delta_{\epsilon_{2}}] \, A_{\mu}^{\cal M} &=& 
(\delta_{\xi} + \delta_{\Lambda})\, A_{\mu}^{\cal M}  
- \frac{4}{7}\, g \left(Z^{\cal M}{}_{\cal PQ}\,{\cal V}^{\cal P}{}_{IK}{\cal V}^{\cal Q}{}_{JK} -
 \theta^{\cal  M}\delta_{IJ} \right)
 \xi_{\mu}^{IJ} 
 \non
 &&{}
 -\xi^\nu\,\Big( 
  {\cal F}_{\mu\nu}^{\cal M}
  + e \,\varepsilon_{\mu\nu\rho}\,{\cal V}^{\cal M}_{\;\;\;\;A}\, {\cal
    P}^{\rho\, A} \Big)
 \;,
 \label{ddA}
\eea
with $ \xi_{\mu}^{IJ}=-\mathrm{i}\,
  \bar\epsilon_{[1}{}^I\gamma^{\mu}\epsilon_{2]}{}^{J}$
and the non-abelian field strength
\bea 
{\cal F}_{\mu\nu}^{\cal M}&=&
2\partial_{[\mu} A_{\nu]}^\mathcal{M}
 + g\, X_{[\mathcal{NP}]}{}^\mathcal{M}
  \,A_\mu^\mathcal{N} A_\nu^\mathcal{P}
  \;.
\eea
In order to arrive at this result,
one needs the explicit expression 
of the intertwining tensor $Z^{\cal M}{}_{\cal PQ}$
which may be obtained after some calculation by plugging 
(\ref{A123}), (\ref{AB}) into (\ref{Z3}):
\begin{eqnarray}
Z^{\cal M}{}_{\cal PQ}\,{\cal V}^{\cal P}{}_{IK}{\cal V}^{\cal Q}{}_{JK}&=& 
- \ft{7}{2}({\cal V}^{\cal M}{}_{IK} A_1^{KJ} + {\cal V}^{\cal M}{}_{JK} A_1^{KI}) 
+ \ft{7}{2}\, {\cal V}^{\cal M}{}_{A} \Gamma^{(I}_{A \dot A} A_2^{J) \dot A} 
\\ \nonumber
&&{}
+ 7 \,{\cal V}^{\cal M}{}_{ K(I} B_{J)K}  - 
\ft12\,\Big({\cal V}^{\cal M}{}_{ KL} B_{KL}  - \ft{15}{4} {\cal V}^{\cal  M}{}_{A} B_A\Big)\,
 \delta_{IJ}
 \;.
\end{eqnarray} 
A priori, the result (\ref{ddA}) differs by its last two terms from the expected 
supersymmetry algebra (\ref{susyalgebra}).
The last term is precisely the duality equation between scalars and vector fields
in three dimensions and signifies the fact that the supersymmetry algebra
closes only modulo the equations of motion~\cite{deWit:2008ta}.
In order to understand the second term in (\ref{ddA}) we recall that
in the gauged theory the vector fields always appear contracted as
$\widehat\Theta_{\cal M\cal N}\,A_\mu^{\cal M}$
or $\theta_{\cal M}\,A_\mu^{\cal M}$.
Under this contraction, the second term in (\ref{ddA}) 
consistently vanishes as a result of the quadratic constraints
(\ref{QQZ}) and (\ref{QD3trace}).\footnote{As in \cite{deWit:2008ta}
one may alternatively absorb this term into additional gauge transformations
related to the further introduction of two-form tensor fields.}

We have thereby established the full set of 
deformed supersymmetry transformation rules
for the general gauged theory in three dimensions.


\subsection{Equations of motion}
\label{subsec:eqm}


We have now all the ingredients to derive the full deformed theory. 
As the gaugings
with local scaling symmetry do no longer admit an action, one must consider
the deformation directly on the level of the equations of motion.
The general gauging is parametrized by an embedding tensor with
components $\Theta_{\cal MN}$, $\theta_{\cal M}$
which defines covariant derivatives according to (\ref{covariant}), (\ref{ansatz}).
For non-vanishing $\theta_{\cal M}$ the gauge group also includes the
generator of the scaling symmetry~(\ref{scaling}).
The embedding tensor defines the scalar field dependent tensors $A_{1,2,3}$ and $B$
which show up in the modified supersymmetry transformation rules (\ref{susyB}), (\ref{susyB})
derived in the last subsection.

In the computation of the supersymmetry algebra (\ref{ddA}), we have
already met the first dynamical equations
\bea
\widehat{\Theta}_{\cal MN}\,
\Big( 
  {\cal F}_{\mu\nu}{}^{\cal M}
  + e \,\varepsilon_{\mu\nu\rho}\,{\cal V}^{\cal M}_{\;\;\;\;A}\, {\cal
    P}^{\rho\, A} \Big)
&=& 0\;,
\nonumber\\
{\theta}_{\cal M}\,
\Big( 
  {\cal F}_{\mu\nu}{}^{\cal M}
  + e \,\varepsilon_{\mu\nu\rho}\,{\cal V}^{\cal M}_{\;\;\;\;A}\, {\cal
    P}^{\rho\, A} \Big)
&=& 0\;.
\label{duality}
\eea
Note that this first order duality equation between vector and scalar fields
is only imposed under projection with $\widehat{\Theta}_{\cal MN}$
and $\theta_{\cal M}$, respectively.
This implies that not the full set of bosonic field equations
but only a projection thereof can be retrieved 
from integrability of this equation. In particular, all contributions from a possible
scalar potential will be invisible in the second order scalar field equations
obtained from~(\ref{duality}).

In order to find the full set of field equations, we
start from the equations of motion of the gravitino of the ungauged theory~\cite{Marcus:1983hb}
\bea
i\gamma^{\rho \mu \nu} D_{\mu} \psi_{\nu}^I  -
 \frac{1}{2} \gamma^{\nu} \gamma^{\rho} \chi^{\dot A} \Gamma ^I _{A \dot A} {P}_{\nu}^A  
 &=& 0
 \;.
\eea
Upon gauging, derivatives are covariantized, i.e.\ $D_\mu \rightarrow {\cal D}_\mu$, 
$P^A_\mu\rightarrow {\cal P}_\mu^A$.
Moreover, in absence of a $\theta_{\cal M}$ the right-hand side of this equation
is modified by terms proportional to the tensors $A_1$ and $A_2$ from 
(\ref{A123})~\cite{Nicolai:2000sc}.
It is thus natural to assume that for the full gauging the r.h.s.\ also receives corrections
proportional to the tensors $B_{IJ}$ and $B_A$. Up to factors $\alpha_{1,2}$, 
these are entirely determined 
by their ${\rm SO}(16)$ structure:
\begin{eqnarray}
i\gamma^{\rho \mu \nu} {\cal D}_{\mu} \psi_{\nu}^I  -
 \frac{1}{2} \gamma^{\nu} \gamma^{\rho} \chi^{\dot A} \Gamma ^I _{A \dot A} {\cal P}_{\nu}^A  
 &=& 
 -  g \left(A_1 ^{IK} +\alpha_1\, B_{IK}\right) \gamma^{\rho \nu} \psi_{\nu}^K
 \nonumber\\[.5ex]
&&{} 
+  i  g \left(A_2 ^{I \dot A} +\alpha_2\,  \Gamma^I _{A \dot A}B_A \right) \gamma^{\rho} \chi^{\dot A}
\;.
\label{eqm:gravitino}
\end{eqnarray}
In order to verify consistency and to determine the factors $\alpha_{1,2}$ we compute
the transformation of this equation under supersymmetry.
We will in this calculation neglect cubic terms in the fermions, i.e.\
only consider variation of the fermionic fields in (\ref{eqm:gravitino}).

The first term gives rise to a contribution involving the commutator
of two covariant derivatives (\ref{covexplicit})
which can be simplified using~(\ref{sugraBianchi}) to
\bea
i\gamma^{\rho \mu \nu}\, {\cal D}_{\mu} {\cal D}_{\nu} \epsilon^I &=&
\ft{i}2
\gamma^{\rho \mu \nu} 
\Big({\cal Q}_{\mu\nu}^{IJ}\, \epsilon^J - \ft12{\cal F}_{\mu\nu} \,\epsilon^I
+\ft14 \widehat{{\cal R}}_{\mu\nu}{}^{ab}\,\gamma_{ab} \,\epsilon^I
\Big)
\nonumber\\
&=&
\ft{i}2
\gamma {\cal Q}_{\mu\nu}^{IJ}\, \epsilon^J 
+\ft{i}2 \Big(
(\widehat{{\cal R}}^{(\rho\mu)}\!-\!\ft12g^{\rho\mu}\,\widehat{{\cal R}})\,\gamma_\mu
- \ft12  {\cal F}_{\mu\nu} \gamma^{\rho \mu \nu}
-\ft12\,{\cal F}^{\rho\mu}\gamma_\mu
\Big)\,\epsilon^I
\;,
\nonumber\\[.5ex]
\label{varg1}
\eea
with the abelian field strength
${\cal F}_{\mu\nu} =
g\theta_{\cal M}\,{\cal F}_{\mu\nu}^{\cal M}$
and
\bea
{\cal Q}_{\mu \nu}^{IJ} 
&\equiv&
2\partial_{[\mu}\,{\cal Q}^{IJ}_{\nu]}
+2{\cal Q}^{K[I}_\mu\,{\cal Q}^{J]K}_\nu
~=~
- \ft{1}{2} \Gamma^{IJ}_{AB} {\cal P}_{\mu}^A {\cal P}_{\nu}^B  
- g {\cal F}_{\mu \nu}^{\cal M} \,\widehat\Theta _{\cal M N} {\cal V}^{\cal N}{}_{IJ}
\;,
\eea
obtained from integrability of (\ref{PQ}).
Likewise, variation of the second term on the l.h.s.\ of~(\ref{eqm:gravitino})
creates terms bilinear in ${\cal P}^A_\mu$ which after some calculation
simplify to
\bea
\ft{i}4\,\gamma^{\rho\mu\nu}\,\Gamma^{IJ}_{AB}{\cal P}_{\mu}^A {\cal P}_{\nu}^B  
\epsilon^J
-\ft{1}2i\Big({\cal P}^{\rho A} {\cal P}^{\mu A}
-\ft12g^{\rho\mu}\,{\cal P}^{\nu A} {\cal P}_\nu^{A}  \Big)\,\gamma_\mu\,\epsilon^I
\;.
\label{varg2}
\eea
The total variation of the l.h.s.\ of~(\ref{eqm:gravitino}) is thus given
by the sum of (\ref{varg1}) and (\ref{varg2}) together with the order $g$
contributions from (\ref{susyF}). Altogether we obtain
\bea
\delta_\epsilon({\rm l.h.s.}) &=&
\ft12i  \Big(
\widehat{{\cal R}}^{(\rho\mu)}-\ft12g^{\rho\mu}\,\widehat{{\cal R}}
-{\cal P}^{\rho A} {\cal P}^{\mu A}
+\ft12g^{\rho\mu}\,{\cal P}^{\nu A} {\cal P}_\nu^{A} \Big)
\gamma_\mu \epsilon^I
\nonumber\\
&&{}
-\ft12 ig
\gamma^{\rho \mu \nu} {\cal F}_{\mu \nu}^{\cal M} \,
\widehat\Theta _{\cal M N} {\cal V}^{\cal N}{}_{IJ}
\, \epsilon^J 
- \ft14i\gamma^{\rho \mu \nu}  {\cal F}_{\mu\nu} \epsilon^I
-\ft14i\,{\cal F}^{\rho\mu}\gamma_\mu \epsilon^I
\nonumber\\
&&{}
-g\gamma^{\rho \mu \nu} {\cal D}_{\mu} 
\Big((A_1 ^{IJ}-B_{IJ}) \gamma_{\nu} \epsilon^J \Big)
 -
\ft12g \gamma^{\nu} \gamma^{\rho} \Gamma ^I _{A \dot A} {\cal P}_{\nu}^A
\left( A_2^{J{\dot A}}+\ft14\, \Gamma^J_{B {\dot A}} B_B \right) \epsilon^J
\;.
\nonumber
\eea
Using the duality equations~(\ref{duality}) to replace the various field strengths
and the differential relations~(\ref{ABderivatives}), this variation reduces to
\bea
\delta_\epsilon({\rm l.h.s.}) &=&
\ft12i  \Big(
\widehat{{\cal R}}^{(\rho\mu)}-\ft12g^{\rho\mu}\,\widehat{{\cal R}}
-{\cal P}^{\rho A} {\cal P}^{\mu A}
+\ft12g^{\rho\mu}\,{\cal P}^{\nu A} {\cal P}_\nu^{A} \Big)
\gamma_\mu \epsilon^I
\nonumber\\
&&{}
+g
\Gamma^{[I}_{A\dot A} A_2^{J]\dot A}
\, {\cal P}^{\rho A} \epsilon^J 
-\ft14 g
\Gamma^{IJ}_{AB}B_B
\, {\cal P}^{\rho A} \epsilon^J 
+ \ft12 g B_A  {\cal P}^{\rho A}\epsilon^I
+\ft14  g \gamma^{\rho\mu} B_A {\cal P}_\mu^{A} \epsilon^I
\nonumber\\
&&{}
-g\gamma^{\rho \mu} 
\Big(\Gamma^{(I}_{A\dot{A}} A_2^{J)\dot{A}}+\ft12\Gamma^{IJ}_{AB}B_B\Big)\, 
{\cal P}_\mu^A\, \epsilon^J
-g\gamma^{\rho \mu \nu} 
(A_1 ^{IJ}-B_{IJ}) \gamma_{\nu}  {\cal D}_{\mu}\epsilon^J
\nonumber\\
&&{}
 +
\ft12g \gamma^{\rho\nu}  \Gamma ^I _{A \dot A}
\left( A_2^{J{\dot A}}+\ft14\, \Gamma^J_{B {\dot A}} B_B \right) 
 {\cal P}_{\nu}^A \epsilon^J
 -
\ft12g  \Gamma ^I _{A \dot A} 
\left( A_2^{J{\dot A}}+\ft14\, \Gamma^J_{B {\dot A}} B_B \right)
{\cal P}^{\rho A} \epsilon^J
\;,
\nonumber
\eea
where we have also made use of the relation
${\cal V}^{\cal M}{}_{\!\! A}\,
\widehat\Theta _{\cal M N} {\cal V}^{\cal N}{}_{\!\!IJ}\,=
\Gamma^{[I}_{A\dot A} A_2^{J]\dot A} \!+\! \ft14 \Gamma^{IJ}_{AB}B_B$,
obtained from (\ref{ansatz}) with (\ref{A123}), (\ref{AB}).
Collecting terms, we finally arrive at
\bea
\delta_\epsilon({\rm l.h.s.}) &=&
\ft12i  \Big(
\widehat{{\cal R}}^{(\rho\mu)}-\ft12g^{\rho\mu}\,\widehat{{\cal R}}
-{\cal P}^{\rho A} {\cal P}^{\mu A}
+\ft12g^{\rho\mu}\,{\cal P}^{\nu A} {\cal P}_\nu^{A} \Big)
\gamma_\mu \epsilon^I
\nonumber\\
&&{}
-\ft18 g \Big(
4
\Gamma^{J}_{A\dot A} A_2^{I\dot A}
+3 \Gamma^{IJ}_{AB}B_B
- 3 \delta^{IJ} B_A  \Big)  
\Big( {\cal P}^{\rho A}+ \gamma^{\rho\mu}{\cal P}_\mu^{A} \Big) 
\,\epsilon^J
\nonumber\\[1ex]
&&{}
-g 
(A_1 ^{IJ}-B_{IJ})\,\gamma^{\rho  \nu}   {\cal D}_{\nu}\epsilon^J
 \;.
 \label{dlhs}
\eea
While the first term will be part of the Einstein equations,
the remaining terms cannot be part of any bosonic equations
of motion and must therefore be cancelled by
the variation of the r.h.s.\ of (\ref{eqm:gravitino}).
As this variation is given by
\bea
\delta_\epsilon({\rm r.h.s.}) &=&
-g 
(A_1 ^{IJ}+\alpha_1\,B_{IJ})\,\gamma^{\rho  \nu}   {\cal D}_{\nu}\epsilon^J
  \nonumber\\[1ex]
&&{}
-\ft12  g  \Gamma^{J}_{A {\dot A}} 
\left(A_2 ^{I \dot A} +\alpha_2\,  \Gamma^I _{B \dot A}B_B \right) 
\gamma^{\rho}  \gamma^{\mu} \, {\cal P}_{\mu}^{A} \epsilon^J 
~+~ {\cal O}(g^2)
\label{drhs}
 \;,
\eea
one observes immediately that with $\alpha_1=-1$, $\alpha_2=-\ft34$,
all terms in order $g$ cancel against (\ref{dlhs}).
It remains to study the order $g^2$ terms in (\ref{drhs}). 
Note that by now we have fixed all 
free parameters, i.e.\ the remaining terms pose a non-trivial
consistency check on the supersymmetry of the equations of motion.
Applying the order $g$ variation on the r.h.s.\
of (\ref{eqm:gravitino}), we obtain
\bea
\delta^{(g^2)}_\epsilon({\rm r.h.s.}) &=&
 - 2 ig^2 \left(A_1 ^{IK} +\alpha_1\, B_{IK}\right) 
 \left(A_1 ^{KJ}-B_{KJ}\right)
 \gamma^{\rho}  \epsilon^J
 \nonumber\\[.5ex]
&&{} 
+  i  g^2 \left(A_2 ^{I \dot A} +\alpha_2\,  \Gamma^I _{A \dot A}B_A \right)
\left( A_2 ^{J {\dot A}}+\ft14 \Gamma^J_{B {\dot A}} B_B \right)  \gamma^{\rho}  \epsilon^J
\;.
\eea
The result can be simplified upon expanding the products and
using the bilinear relations
between the tensors $A_{1,2,3}$, $B$, derived in 
section~\ref{subsec:quadratic}.
We first observe that the combination $2A_1A_1-A_2A_2$
can be replaced using (\ref{4-1}).
Furthermore, by virtue of (\ref{3-3}) and (\ref{3-32}) 
we can eliminate all the
$\Gamma^I B_A A_2$ terms and obtain altogether
\bea
\delta^{(g^2)}_\epsilon({\rm r.h.s.}) &=&
-ig^2 
\Big(
(2(\alpha_1-\alpha_2)+\ft12) \, B_{IK}B_{JK}
+(2\alpha_1-3\alpha_2-\ft14)\, B_{IK} A_1^{JK} 
\nonumber\\
&&{}
+(\alpha_2+\ft34)\,B_{JK} A_1^{IK}
-\ft1{16}\,\delta^{IJ} 
(2 A_1^{KL}A_1^{KL} -A_2^{K\dot A}A_2^{K\dot A}-B_AB_A)
\Big) \, 
 \gamma^{\rho}  \epsilon^J
 \;.
 \nonumber\\
 \label{drfinal}
\eea
Remarkably, with the choice $\alpha_1=-1$, $\alpha_2=-\ft34$
imposed earlier, the first three terms in this variation vanish
and the result is again proportional to $\delta^{IJ}$
and can thus be absorbed into the Einstein equations, 
as required for consistency.
Combining (\ref{drfinal}) with (\ref{dlhs}) and (\ref{drhs})
we thus finally obtain the modified Einstein equation
\bea
\widehat{{\cal R}}_{(\mu\nu)}-\ft12g_{\mu\nu}\,\widehat{{\cal R}}
&=&
{\cal P}^A_{\mu} {\cal P}^A_{\nu}
-\ft12g_{\mu\nu}\,{\cal P}^{\rho A} {\cal P}_\rho^{A}
-\ft12 g_{\mu\nu}\,W(\phi)
\;,
\label{eqmEinstein}
\eea
with
\bea
W(\phi) &\equiv& 
\ft14 g^2
\Big(A_2^{I\dot A}A_2^{I\dot A}+B_AB_A-2 A_1^{IJ}A_1^{IJ} \Big)
\;,
\label{potentialnew}
\eea
playing the role of an effective 
(scalar field dependent) cosmological constant in this equation.
Comparing this result to (\ref{potentialold}),
we observe that the effect of a gauging of the scaling symmetry $\mathbb{R}^+$
is a positive contribution to this effective cosmological constant.
The same effect will occur in the corresponding higher-dimensional theories.
In standard gravity theories the scalar dependent function $W(\phi)$
would correspond to the scalar potential from which in particular also the
scalar masses are derived.
This is different in the presence of an $\mathbb{R}^+$-gauging:
as the resulting theory does in general not admit an action,
it is not clear if the mass contributions to the scalar field equations
descend from a scalar potential --- and we will see 
in equation~(\ref{eqmscalar}) below explicitly
that this is not the case. To this end, we note that the variation of~(\ref{potentialnew})
is given by
\bea
\delta\, W(\phi) &=& 
-\ft14g^2 \Gamma^I_{A\dot{A}} \Big(
3 A_1^{IJ}A_2^{J\dot A} 
- A_2^{I\dot B} A_3^{\dot{A}\dot{B}}
+\ft1{8} \Gamma^{J}_{B\dot{A}} B_{IJ} B_B 
\Big)\;\delta\Sigma^A
\;,
\label{potentialvar}
\eea
as can be derived from the differential relations~(\ref{ABderivatives})
upon replacing ${\cal P}^A_{\!\mu}\,$ by $\delta\Sigma^A$.

By calculating the supersymmetry variation of the gravitino field equation
we have thus fixed all unknown coefficients in this equation and
obtained the modified Einstein equation up to its fermionic contributions.
The latter may in principle be obtained by repeating the calculation
including all higher order fermionic terms.
The remaining set of equations of motion are the Dirac equation for
the spin-$1/2$ fields and the scalar field equation. These may be 
determined in complete analogy to the calculation presented. 
Rather than going once more through the technical details, we just 
present the resulting equations:
\begin{eqnarray}
\gamma^{\mu} {\cal D}_{\mu} \chi^{\dot A}  &=& 
\ft{i}{2} \gamma^{\mu} \gamma^{\nu} \psi_{\mu}^I 
\Gamma^I _{A \dot A} {\cal P}_{\nu}^A
+
g \left( A_2^{I \dot A} + \ft{1}{4}  \Gamma^I_{A \dot A} B_A \right) \gamma^{\mu} \psi_{\mu}^I 
- i g \left( A_3^{\dot A \dot B} \!+ \ft{1}{4}  B_{IJ} \Gamma^{IJ}_{\dot A \dot B} \right) \chi^{\dot B} 
\;,
\nonumber\\
{\cal D}^{\mu} {\cal P}_\mu^{A} &=& 
\ft{1}{8} \, g^2 \, \Gamma^I_{A \dot A} \left( 3 \, A_1^{IJ} A_2^{J \dot A} - A_3^{\dot A \dot B} A_2^{I \dot B} + 2 \, B_{IJ} A_2^{J \dot A} - 2 \, \Gamma^J_{B \dot A} B_{IJ} B_B \right)
\;.
\label{eqmscalar}
\eea
The quadratic constraint~(\ref{final2}) crucially enters in the derivation of
these equations.
We have thus obtained the full set of deformed equations
of motion for the general gauged maximal theory in three dimensions
to lowest order in the fermions.
Comparing (\ref{eqmscalar}) to (\ref{potentialvar}) one observes that
the scalar mass terms (the r.h.s.\ of~(\ref{eqmscalar})) for non-vanishing $B_A$
do not descend from the potential $W(\phi)$.
This is another manifestation of the fact that the resulting theory does not admit an action.


\section{Conclusions and Outlook}


In this paper we have constructed the gaugings of
maximal supergravity in which the trombone symmetry~(\ref{scaling})
becomes part of the local gauge symmetries.
We have set up the algebraic formalism to describe these theories
as an extension of the standard gaugings. 
More precisely, the gaugings are parametrized 
by a constant embedding tensor $\widehat{\Theta}_M{}^{\hat\alpha}$ which has
irreducible components $(\Theta_M{}^\alpha, \theta_M)$.
In case the second component is zero, $\theta_M=0$, 
these theories reduce to the standard gaugings
with gauge group inside the duality group~${\rm G}$. 
Non-vanishing $\theta_M$ on the other hand amounts to 
the inclusion of the scaling symmetry~(\ref{scaling})
into the gauge group.

The explicit form of the gauge group generators is
given in~(\ref{ansatz}) where the value of $\zeta$ has been 
determined in section~\ref{sec:algebraic} for the 
maximal supergravities in various dimensions.
As a result we find that gauging of the scaling symmetry~(\ref{scaling})
necessitates simultaneous gauging of certain generators
within the duality group~${\rm G}$.
We have worked out the algebraic consistency constraints
bilinear in the components $(\Theta_M{}^\alpha, \theta_M)$.
For the particular class of theories with $\Theta_M{}^\alpha=0$
(which thus correspond to a ``minimal'' gauging of the trombone symmetry),
we have explicitly constructed the general solution to these
consistency constraints. Interestingly, this solution relies on
a generalization of the ``pure spinor'' structure of ${\rm SO}(10)$
to the higher-rank exceptional groups.

Finally, we have for the example of the three-dimensional theory
worked out the deformed supersymmetry algebra and the full 
set of equations of motion.
In particular, we have shown that gaugings involving the
trombone symmetry are compatible with supersymmetry
provided the components of the embedding tensor satisfy the
aforementioned algebraic consistency constraints. 
Since these theories in general do no longer admit an action
they must be constructed on the level of the equations of motion
which are uniquely determined by supersymmetry.

As a generic feature of a gauging of the trombone symmetry
we have found a positive contribution to the cosmological constant.
The same shows up in the corresponding higher-dimensional theories.
The existence of a ten-dimensional de Sitter vacuum in the 
theory of~\cite{Howe:1997qt,Lavrinenko:1997qa} has been 
further investigated in~\cite{Chamblin:2001dx}.
From this point of view it will be interesting to analyze the 
general structure of the equations of motion
and their solutions for the theories with ``minimal'' 
gauging of the trombone symmetry given in this paper.
Another interesting question is about the structure of theories for which
both components $\Theta_M{}^\alpha$ and $\theta_M$ are non-vanishing.
The presence of additional deformation parameters $\theta_M$
as compared to the standard gaugings (which moreover give rise to 
positive contributions in the cosmological constant) may prove useful
in the search for stable de Sitter vacua in 
$N>1$ supersymmetric theories which to date seem 
extremely rare~\cite{Fre:2002pd,deRoo:2003rm}.
Of course, a higher-dimensional interpretation for these
additional deformation parameters would be highly desirable.
\smallskip

Let us finally discuss another intriguing aspect about the theories 
we have constructed. It is well known
that the representation in which the embedding tensor
transforms under ${\rm G}$ in the standard gaugings 
(column `$\Theta$' of table~\ref{tabTT})
is the representation dual to the 
totally antisymmetric $(D-1)$-forms of the theory as
predicted from the underlying very extended
Kac-Moody algebra ${\rm E}_{11}$~\cite{Riccioni:2007au,Bergshoeff:2007qi}.
More precisely, the embedding tensor can be identified with the
integration constants which arise upon solving the non-dynamical
field equations for the $(D-1)$-forms~\cite{deWit:2008ta}.
In contrast, the additional gaugings we have constructed allow
for additional components $\theta_M$ of the embedding tensor
transforming in the representation dual to the vector fields.
For these constants there is no dual $(D-1)$-form in the field content
of the theories, i.e.\ an ${\rm E}_{11}$ origin of these theories
is a priori unclear. 
However, following the discussion in the introduction, the
trombone symmetries in the various dimensions seem intimately
linked to the duality groups ${\rm G}$, such that one would
expect that all these gaugings can be cast into a common framework.
Indeed, some observations hint in this direction: 
inspecting a little closer the full field content as predicted 
by ${\rm E}_{11}$, as given in the tables of~\cite{Bergshoeff:2007qi},
one observes that there does exist an object in the correct ${\rm G}$
representation with $D-1$ space-time indices which however is
not an antisymmetric form but
a tensor with mixed symmetry $C^M_{\mu,[\nu_1\nu_2 \dots , \nu_{D-2}]}$.
Like the $(D-1)$-forms, such a field 
does not possess propagating degrees of freedom
(see e.g.~\cite{Bekaert:2002dt}) and can consistently be set to zero.
It is a highly intriguing question if the presence of such tensors
with mixed symmetry could in some way 
trigger the deformations of the presented type.

In fact, the pattern continues: the antisymmetric $D$-forms in standard
gaugings turn out to transform under ${\rm G}$ in the representation
which is dual to the quadratic constraint on the embedding tensor~\cite{deWit:2008ta}.
As we have shown in this paper, the presence of the additional
components $\theta_M$ gives rise to additional quadratic constraints,
cf.~equations~(\ref{qreps6}), (\ref{qreps5}), (\ref{qreps4}), (\ref{qreps3}),
for the various dimensions. Comparing these additional representations
to the tables of~\cite{Bergshoeff:2007qi} we find again a matching of
representations with tensors carrying $D$ space-time indices with mixed 
symmetry structure! 

After reduction to $D=2$ dimensions, all these tensors embed into representations
of the affine symmetry algebra ${\rm E}_{9(9)}$.
Table~\ref{tabTT} shows that remarkably under this algebra there is no longer a 
difference between the theories triggered by
the new parameters $\theta$ and the standard gaugings: both $\Theta$
and $\theta$ combine into a single irreducible 
(infinite-dimensional) representation of ${\rm E}_{9(9)}$~\cite{Samtleben:2007an},
suggesting that also under the bigger algebras 
${\rm E}_{10}$ and ${\rm E}_{11}$ there should be a uniform 
and common structure underlying all the gaugings.

\smallskip

Along these lines, let us recall that as we have seen throughout 
the construction, gaugings that involve a local trombone symmetry 
do no longer admit an action and have thus been constructed on
the level of the equations of motion. This is by no means surprising since
they involve the gauging of a symmetry that was not off-shell realized.
However, a similar fate applies to part of the duality groups ${\rm G}$
in even space-time dimensions. E.g.\ in $D=4$ dimensions 
(depending on the electric frame chosen) only
an ${\rm SL}(8)$ subgroup of ${\rm G}={\rm E}_{7(7)}$ is realized as
a symmetry of the action while the full ${\rm E}_{7(7)}$ can only be
realized on the combined set of equations of motion
and Bianchi identities~\cite{Cremmer:1979up}.
Nevertheless, in this theory it is possible to gauge subgroups
within the full ${\rm E}_{7(7)}$ on the level of the action ---
upon introducing further higher-rank $p$-forms~\cite{deWit:2005ub}.
The same pattern extends to all even 
dimensions~\cite{deWit:2007mt,Samtleben:2007an,Bergshoeff:2007ef}.
It would be very exciting (and further complete the presumed ${\rm E}_{11}$
picture underlying the theory) if also the theories presented in this paper could be lifted
to an action precisely by introducing the additional higher-rank tensors
of mixed symmetry mentioned above.
In this respect we mention the recent construction of a
parent action for the dual graviton --- the simplest of all
tensors with mixed symmetry --- which is based on St\"uckelberg-type
couplings to higher-rank tensor fields in a way reminiscent of the 
structures appearing in gauged supergravity~\cite{Boulanger:2008nd}.

\bigskip
\bigskip


\noindent
{\bf Acknowledgements:}

This work is supported in part by the Agence Nationale de la Recherche (ANR).


\bigskip
\bigskip

\appendix

\section*{Appendix}


\section{Algebra conventions}



\subsection{${\rm E}_{8(8)}$ conventions.}
\label{app:e8}


The algebra $\mathfrak{e}_{8(8)}$ is generated by 248 generators $t_{\cal M}$
\bea
{}
[\,t_{\cal M},t_{\cal N}] &=& f_{\cal MN}{}^{\cal K}\,t_{\cal K}
\;,
\eea 
which may be split into 120 compact ones $X^{IJ} = - X^{JI}$, 
corresponding to the maximal compact subalgebra $\mathfrak{so}(16)$ of the algebra, 
and 128 non-compact ones $Y^A$, 
with ${\rm SO}(16)$ vector indices $I,J,... = 1,...,16$, and spinor indices $A,B,... = 1,...,128$. 
Dotted indices $\dot A, \dot B,...$ label the conjugate ${\rm SO}(16)$ spinor representation. 
An extra factor of~$\frac{1}{2}$ always appears when 
summing over antisymmetrized index pairs $[IJ]$.
${\rm E}_{8(8)}$ indices are raised and lowered by means of the
Cartan-Killing metric
\bea
\eta_{\cal MN}&=&\frac1{60} {\rm Tr} \, t_{\cal M} t_{\cal N} 
\;.
\eea
In the ${\rm SO}(16)$ basis, the components of the Cartan-Killing form are
$\eta^{AB}=\delta^{AB}$ and $\eta^{IJ,KL}=-2\delta^{IJ}_{KL}$
and the completely antisymmetric 
structure constants of the algebra are 
given by
\bea
f^{IJ,\, KL,\, MN} &=& 
-8 \, \delta^{[I[K} \delta^{L]J]}_{MN} \;,\qquad 
f^{IJ,\, A,\, B} = -\frac{1}{2} \, \Gamma^{IJ}_{AB}
\;.
\label{structure}
\eea
An important object is the group-valued scalar matrix 
${\cal V}^{\cal M}{}_{\underline{\cal M}} =\{ {\cal V}^{\cal M}{}_{IJ}, {\cal V}^{\cal M}{}_{A} \}$.
It satisfies
\bea
\eta_{\cal MN}\,
{\cal V}^{\cal M}{}_{IJ}
{\cal V}^{\cal N}{}_{KL}
=
-2\delta^{IJ}_{KL}
\;,
\qquad
\eta_{\cal MN}\,
{\cal V}^{\cal M}{}_{A}
{\cal V}^{\cal N}{}_{B}
=
\delta_{AB}
\;,
\eea
which allows to express its inverse explicitly as
\bea
{\cal V}^{\underline{\cal M}}{}_{\cal M} &=&
\left\{
\begin{array}{rcrcl}
{\cal V}^{IJ}{}_{\cal M} &=& 
-{\cal V}_{\cal M}{}^{IJ}\!\!\!&\equiv&\!\! \eta_{\cal MN}{\cal V}^{{\cal N} IJ}\\
{\cal V}^{A}{}_{\cal M} &=& 
{\cal V}_{\cal M}{}^{A}\!\!\!&\equiv&\!\! \eta_{\cal MN}{\cal V}^{{\cal N} A}
\end{array}
\right.
\;.
\eea
The fact that the structure constants~(\ref{structure})
are ${\rm E}_{8(8)}$ invariant tensors and thus
invariant under contraction with ${\cal V}^{\cal M}{}_{\underline{\cal M}}$ 
is reflected by equations~(\ref{fff}).


\subsection{${\rm E}_{7(7)}$ conventions and identities.}
\label{app:e7}


The algebra $\mathfrak{e}_{7(7)}$ is generated by 133 generators $t_\alpha$
\bea
[t_{\alpha},\, t_{\beta}] &=& f_{\alpha \beta}{}^{\gamma}\, t_{\gamma}
\;.
\eea
Its fundamental representation has dimension 56; indices $m, n = 1, \dots, 56,$
can be raised and lowered with the symplectic matrix $\Omega_{mn}$
where we use north-west south-east conventions
\bea
X^m &=& \Omega^{mn} X_n \;, \qquad  X_m = X^n \Omega_{nm}
\;.
\eea
We raise and lower the adjoint indices $\alpha, \beta = 1, \dots 133,$ with
the invariant metric $\kappa^{\alpha \beta} = \textrm{Tr}(t^{\alpha} t^{\beta})$
proportional to the Cartan-Killing form.
It is related to the structure constants $f_{\alpha \beta}{}^{\gamma}$ as
\bea
f_{\alpha\gamma}{}^{\delta} f_{\beta\delta}{}^{\gamma} =  3\, \kappa_{\gamma \delta}
\;.
\eea

By performing various contractions, one can prove 
the non-trivial relation~(\ref{relD4}) between $E_{7(7)}$ generators
\bea
(t^\alpha)_m{}^{k} (t_\alpha)_n{}^{l} &=&
\ft1{24}\delta_m^{k}\delta_n^{l} + \ft1{12}\delta_m^{l}\delta_n^{k} 
+(t^{\alpha})_{mn}\,(t_{\alpha})^{kl}-\ft1{24}\Omega_{mn}\,\Omega^{kl}
\;.
\eea
E.g., contracting the indices $k$ and $n$, we find in particular
\bea
8\,(t^{\alpha})_{m}{}^{k} (t_{\alpha})_{k}{}^{l} &=& 19 \,\delta_m^l
\;.
\eea

We will need some more identities for this algebra. The first one takes the form
\bea 
9 (t^{\alpha})_m{}^{k} (t^{\beta})_{kn} (t_{\alpha})^{(pq} (t_{\beta})^{rs)} 
+2(t^{\alpha})^{\vphantom{M}}_{[m}{}_{\vphantom{M}}^{(r} (t_{\alpha})_{\vphantom{[}}^{\vphantom{[}pq} \delta_{n]}^{s)} 
&=& \ft{1}{8} \, \Omega_{mn} (t^{\alpha})^{(pq} (t_{\alpha})^{rs)} 
\;.
\label{identity1}
\eea
Note that this identity is antisymmetric in $[mn]$ and totally symmetric in $(pqrs)$.
The existence of such a relation thus follows from the fact that there are only two
independent invariant tensors with this index structure (only two singlets in the
correspondingly symmetrized tensor product of fundamental representations). The coefficients can be determined by performing various contractions.

In a similar way we obtain another important relation 
which is totally symmetric in indices $(klmnpq)$:
\bea
 (t^{\beta})^{(kl} (t_{\beta})^{mn} (t_{\alpha})^{pq)}  
+8 
(t_{\alpha})^{rs} (t^{\beta})_r{}^{(k} (t^{\gamma})_s{}^{l}  
(t_{\beta})^{mn} (t_{\gamma})^{pq)}
&=&0
\;.
\label{identity2}
\eea


\subsection{${\rm E}_{8(8)}$ algebra in the ${\rm E}_{7(7)} \times {\rm SL}(2)$ basis}
\label{subsec:e8e7}


Under its maximal subgroup ${\rm E}_{7(7)} \times {\rm SL}(2)$, the adjoint representation
of ${\rm E}_{8(8)}$ breaks as
\bea
{\bf 248} &\longrightarrow&
({\bf 133},{\bf 1}) \oplus  ({\bf 56},{\bf 2}) \oplus ({\bf 1},{\bf 3}) 
\;.
\label{break248}
\eea
Accordingly, we split the generators $t_{\cal M}$ into $t_{\alpha}$, $t_{m,a}$ and $t_{(ab)}$,
where $m$ and $\alpha$ denote the fundamental and the adjoint representation
of ${\rm E}_{7(7)}$, respectively, while $a, b$ denote the doublet of ${\rm SL}(2)$.
In these generators, the algebra takes the form
\bea
{}[t_{\alpha},\, t_{\beta}] &=& f_{\alpha \beta}{}^{\delta} \, t_{\delta} \;,
\qquad
[t_{(ab)},\,  t_{(cd)}] ~=~ 2 \delta_{(a}^{e\vphantom{f}} \, \epsilon^{\vphantom{f}}_{b)(c} \, \delta_{d)}^{f} \, t_{(ef)}
\;,
\nonumber\\[.5ex]
{}[t_{m,\, a},\, t_{(bc)}] &=& \epsilon_{a(b} \, t_{m,\, c)} 
\;,\qquad 
[t_{m,\, a},\, t_{\alpha}] ~=~ (t_{\alpha})_{m}{}^{n} \, t_{n,\, a}
\;,
\nonumber\\[.5ex]
{}[t_{m,\, a},\, t_{n,\, b}] &=& \ft{1}{12} 
\Omega_{mn} \, t_{(ab)} + \epsilon_{ab} \, (t^{\alpha})_{mn} \, t_{\alpha}
\;.
\label{e8e7algebra}
\eea
Here, we use the ${\rm E}_{7(7)}$-invariant tensors introduced in section~\ref{app:e7}
and the ${\rm SL}(2)$-invariant $\epsilon$-symbol $\epsilon^{ab}$.

\bigskip


\section{Solution of the ${\rm E}_{8(8)}$ constraint~(\ref{pure3})}
\label{app:constraint}


A particular class of gaugings we have studied in this paper
are those theories which are triggered by a single constant vector $\theta_M$.
In three dimensions, this vector is subject to the quadratic constraint~(\ref{pure3})
\bea
\Big(
({\mathbb P}_{\bf 1})_{\cal MN}{}^{\cal KL} +
({\mathbb P}_{\bf 3875})_{\cal MN}{}^{\cal KL}
\Big)\;\theta_{\cal K}\,\theta_{\cal L} &\stackrel!{\equiv}& 0
\;.
\label{pureA}
\eea
In this appendix we will analyze in detail this quadratic constraint
and derive its general solution given in~(\ref{sol3}) in the main text.
Explicitly, the constraint~(\ref{pureA}) reads
\bea
\theta^{\cal  M} \theta_{\cal  M} &=& 0
\;,\qquad
 \theta_{\cal  M} \theta_{\cal N} - \ft{1}{2} f^{\cal  QP}{}_{\cal  M}f_{\cal  NQ}{}^{\cal  L} \,\theta_{\cal  P} \theta_{\cal  L} ~=~ 0
 \;.
\eea


\subsection{The constraint under ${\rm E}_{7(7)}\times {\rm SL}(2)$}
\label{subsec:cone7}


In order to solve the constraint~(\ref{pureA}), we employ the same technique that 
allowed to explicitly solve the pure spinor constraint~(\ref{pure6})
in $D=6$ and its analogues (\ref{pure5}), (\ref{pure4}) in
$D=5$ and $D=4$, respectively.
It is useful to first break ${\rm E}_{8(8)}$ under its
subgroup ${\rm E}_{7(7)}\times {\rm SL}(2)$ as 
given explicitly in section~\ref{subsec:e8e7} above. 
The adjoint representation breaks according to (\ref{break248})
such that we can parametrize the 
vector $\theta_{\cal M}$ by components
$\{ \theta_\alpha, \theta_{m,a}, \theta_{(ab)} \}$.
The constraint~(\ref{pureA}) under this subgroup breaks into
\bea
\mathbf{1} \oplus \mathbf{3875} &\longrightarrow& 
2\cdot\mathbf{(1,\, 1)} \oplus \mathbf{(1539, 1)} \oplus 
\mathbf{(56, 2)} \oplus  \mathbf{(912, 2)} \oplus  \mathbf{(133,\, 3)}
\;. 
\label{break3875}
\eea
As a first step we will express this constraint explicitly in terms of 
the ${\rm E}_{7(7)}\times {\rm SL}(2)$ components $\{ \theta_\alpha, \theta_{m,a}, \theta_{(ab)} \}$.
To this end, we start from the general ${\rm E}_{7(7)}\times {\rm SL}(2)$ singlet
bilinear in the components of $\theta_{\cal M}$
\bea
\Phi_{\sigma,\tau}&\equiv & 
\theta^{(cd)} \, \theta_{(cd)} + 
\sigma \, \epsilon^{cd} \, \Omega^{kp} \, \theta_{k,\, c} \, \theta_{p,\, d} 
 +  \tau \, \theta_{\alpha}  \, \theta^{\alpha} 
 \;,
 \label{singlets}
\eea
labeled by two relative coefficients $\sigma$, $\tau$. 
With respect to ${\rm E}_{8(8)}$, 
the general bilinear expression in $\theta_{\cal M}$
transforms in the representation
\bea
(\mathbf{248} \otimes \mathbf{248})_{\rm sym} &=& 
\mathbf{1} \oplus \mathbf{3875} \oplus \mathbf{27000}
\;.
\eea
In order to identify the constraint~(\ref{pureA}) we seek
within the three linearly independent singlets~(\ref{singlets}) 
the two combinations corresponding to the r.h.s.\ of~(\ref{break3875}),
i.e.\ the two singlets descending from the ${\rm E}_{8(8)}$
representations $\mathbf{1}$ and $\mathbf{3875}$
 --- while the third combination
corresponds to the singlet
descending from the ${\bf 27000}$.

To this end we compute the action of an ${\rm E}_{8(8)}$ generator $t_{m,\, a}$ on $\Phi_{\sigma,\tau}$. As $\theta_{\cal M}$ transforms in the adjoint representation of ${\rm E}_{8(8)}$, this action can directly be deduced from~(\ref{e8e7algebra}).
The result is given by
\bea
\label{adj1}
t_{m, a} \cdot \Phi_{\sigma,\tau} &=& 
\ft16 \, \left( \sigma-12 \right) \, \epsilon^{cd} \, \theta_{m,\,c} \, \theta_{(ad)}  
+  2 \, (\sigma +   \tau) \, (t^{\alpha})_{m}{}^{n} \, \theta_{n,\,a} \, \theta_{\alpha}
\;.
\eea
This shows that $\Phi_{12,-12}$ is an ${\rm E}_{8(8)}$ singlet, i.e.\ we have identified
the singlet descending from the ${\bf 1}$ of ${\rm E}_{8(8)}$.
Applying another generator $t_{n,\, b}$ on (\ref{adj1}) and contracting all
free indices gives rise to
\bea
 \Omega^{nm} \epsilon^{ab}\, t_{n,\, b} \cdot t_{m,\, a} \cdot 
 \Phi_{\sigma,\tau} &=& 
 \ft{7}{9}  \left(12-\sigma \right)   \theta^{(ab)} \, \theta_{(ab)} 
  +4 \, (\sigma  +  \tau)  \theta^{\alpha}  \theta_{\alpha} 
\nonumber\\[.5ex]
&& -\ft14  \left(
12-20\sigma-19\tau \right) \Omega^{mn} \epsilon^{ab} \, \theta_{m,\, a} \, \theta_{n,\, b} 
\;.
\label{actionCasimir}
\eea
Note that the operator ${\cal C}\equiv  \Omega^{nm} \epsilon^{ab}\, t_{n,\, b}  t_{m,\, a}$ acting on $\Phi_{\sigma,\tau}$ is proportional to the quadratic Casimir of ${\rm E}_{8(8)}$.
Diagonalizing its action~(\ref{actionCasimir}) we find
\bea
{\cal C} \cdot \Phi_{12,-12} = 0
\;,\qquad
{\cal C} \cdot \Phi_{\frac{12}7,\frac{12}7} ~=~ 8\,\Phi_{\frac{12}7,\frac{12}7}
\;,\qquad
{\cal C} \cdot \Phi_{-\frac{9}7,-\frac{108}{133}} ~=~ \frac{31}3\, \Phi_{-\frac{9}7,-\frac{108}{133}}
\;,
\eea
and can thereby identify the singlets $\Phi_{12,-12}$ and 
$\Phi_{\frac{12}7,\frac{12}7}$ descending from 
the ${\bf 1}$ and the ${\bf 3875}$ of ${\rm E}_{8(8)}$, respectively.

The full constraint~(\ref{pureA}) can thus be obtained as the 
${\rm E}_{8(8)}$ orbit of the second singlet. The action of two 
${\rm E}_{8(8)}$ generators on $\Phi_{\frac{12}7,\frac{12}7}$
is given by
\bea \label{tt}
7\,t_{n,\, b} \cdot t_{m,\, a} \cdot \Phi_{\frac{12}7,\frac{12}7} &=& -  \Omega_{nm} \, \epsilon^{cd} \, \theta_{(bc)} \, \theta_{(ad)} +  (t^{\alpha})_{nm} \, \theta_{\alpha} \theta_{(ab)} \\ \nonumber
&&{}-6\, \epsilon_{ba} \, \epsilon^{cd} \, \theta_{m,\,c} \, \theta_{n,d} -6\, \theta_{m,\,b} \, \theta_{n,\,a}\\ \nonumber
&&{}+ 48 \, \epsilon_{ba} \, (t^{\alpha})_m{}^k \, (t_{\beta})_{kn} \, \theta_{\beta} \, \theta_{\alpha} +48 \, (t^{\alpha})_m{}^k \, (t_{\beta})_{n}{}^p \, \theta_{k,\, a} \, \theta_{p,\,b}\;.
\eea
By various contractions one finds from this equation and from (\ref{adj1}) 
the different parts of~(\ref{break3875}). 
As a result, we give the constraint~(\ref{pureA}) explicitly in terms of
the components $\{ \theta_\alpha, \theta_{m,a}, \theta_{(ab)} \}$:
\bea
\theta^{(cd)} \, \theta_{(cd)} + 
12 \, \epsilon^{cd} \, \Omega^{kp} \, \theta_{k,\, c} \, \theta_{p,\, d} 
 -12 \, \theta_{\alpha}  \, \theta^{\alpha} &\stackrel!{\equiv}& 0 
 \;, \quad
 (\mathbf{1},\mathbf{1} )_{(1)} 
 \nonumber\\
 7\theta^{(cd)} \, \theta_{(cd)} + 
12 \, \epsilon^{cd} \, \Omega^{kp} \, \theta_{k,\, c} \, \theta_{p,\, d} 
 +  12 \, \theta_{\alpha}  \, \theta^{\alpha} &\stackrel!{\equiv}& 0 
 \;, \quad
 (\mathbf{1},\mathbf{1} )_{(3875)} 
  \nonumber\\
 \epsilon^{cd} \, \theta_{m,\,c} \, \theta_{(ad)}  
-4 (t^{\alpha})_{m}{}^{n} \, \theta_{n,\,a} \, \theta_{\alpha}
   &\stackrel!{\equiv}& 0 
 \;, \quad
 (\mathbf{56},\mathbf{2} )
  \nonumber\\
   \theta_{\alpha} \, \theta_{(ab)} 
  +6 \, (t_{\alpha})^{mn} \, \theta_{m,\, a} \, \theta_{n,\, b} 
     &\stackrel!{\equiv}& 0 
 \;, \quad
 (\mathbf{133},\mathbf{3} )
  \nonumber\\
\epsilon^{ab} \, \theta_{[m,\, a} \, \theta_{n],\, b} - 6 \, (t^{\alpha})_{[m}{}^{k} \, (t^{\beta})_{n]k}  \, \theta_{\alpha} \, \theta_{\beta}
 ~-~ {\rm trace}
     &\stackrel!{\equiv}& 0 
 \;. \quad
 (\mathbf{1539},\mathbf{1} )
 \label{pureB}
\eea
We have left out the constraint in the $({\bf 912},{\bf 2})$ which 
is obtained by the action of three ${\rm E}_{8(8)}$ generators 
on $\Phi_{\frac{12}7,\frac{12}7}$. As we shall see in the next section, 
this part of the constraint is automatically satisfied and does not lead to new constraints.


\subsection{Solving the constraint}


For the explicit solution of~(\ref{pureB}), we further break these
equations under ${\rm E}_{7(7)}\times \mathbb{R}^+$.
According to the decomposition~(\ref{break248R}), we break 
the vector $\theta_{\cal M}$ into components 
$(\tilde\eta, \tilde\eta_m, \xi, \xi_{\alpha}, \eta_m, \eta)$ defined as
\bea
\tilde\eta&\equiv&\theta_{(++)}\;,\quad
\xi ~\equiv~\theta_{(+-)}\;,\quad
\eta~\equiv~\theta_{(--)}\;,
\nonumber\\
\tilde\eta_m &\equiv& \theta_{m,\, +}\;,\quad
\eta_m ~\equiv~ \theta_{m,\, -}\;,\quad
\xi_{\alpha} ~\equiv~\theta_{\alpha}
\;,
\eea
where we have broken up the ${\rm SL}(2)$ components introduced in the last
subsection.
In terms of these components, the full set of constraints~(\ref{pureB})
takes the form
\bea
\eta \, \tilde\eta  -  \xi ^2 +  12 \, \Omega^{mn} \, \tilde\eta_{m} \, \eta_{n}  
-  6 \, \xi_{\alpha} \, \xi^{\alpha}
 &\stackrel!{\equiv}& 0 \;, \quad
 (\mathbf{1}^0_{(1)} ) 
\nonumber\\
7\eta \, \tilde\eta  - 7 \xi ^2  + 
 12 \, \Omega^{mn} \, \tilde\eta_{m} \, \eta_{n}  + 
  6 \, \xi_{\alpha} \, \xi^{\alpha}
 &\stackrel!{\equiv}& 0 \;,\quad
 (  \mathbf{1}^0_{(3875)})
\nonumber\\
\tilde\eta \, \eta_m - \xi \tilde\eta_m  +  
4 \, (t^{\alpha})_{m}{}^{n} \, \xi_{\alpha} \, \tilde\eta_n 
 &\stackrel!{\equiv}& 0 \;,\quad
 (\mathbf{56}^{+1})
\nonumber\\
 \eta \, \tilde\eta_m  -  \xi \eta_m  -  
 4 \, (t^{\alpha})_{m}{}^{n} \, \xi_{\alpha} \, \eta_n 
 &\stackrel!{\equiv}& 0 \;,\quad
 (\mathbf{56}^{-1})
\nonumber\\
 \xi_{\alpha} \, \tilde\eta  +  6 \, (t_{\alpha})^{mn} \, \tilde\eta_m \, \tilde\eta_n 
 &\stackrel!{\equiv}& 0\;, \quad
 (\mathbf{133}^{+2})
\nonumber\\
 \xi_{\alpha} \, \xi  +  6 \, (t_{\alpha})^{mn} \, \tilde\eta_m \, \eta_n 
&\stackrel!{\equiv}& 0\;, \quad
 (\mathbf{133}^{0})
\nonumber\\
 \xi_{\alpha} \, \eta  +  6 \, (t_{\alpha})^{mn} \, \eta_m \, \eta_n 
&\stackrel!{\equiv}& 0 \;, \quad
 (\mathbf{133}^{-2})
 \nonumber\\
 \tilde\eta_{[m} \, \eta_{n]}  
 -  \ft{1}{56}  \Omega_{mn}  \Omega^{kp}  \tilde\eta_k  \eta_p 
 - 3  (t^{\alpha})_{[m}{}^{k} \, (t^{\beta})_{n]k} \xi_{\beta} \xi_{\alpha}  
 +  \ft{3}{56}  \Omega_{mn} \xi^{\alpha}  \xi_{\alpha} 
 &\stackrel!{\equiv}& 0 \;, \quad
 (\mathbf{1539}^{0})
 \nonumber\\
\label{pure0}
\eea
where again we have left out the two equations in the ${\bf 912}^{\pm1}$ 
which we will justify shortly.

In analogy to the higher-dimensional cases, we start from a given
set of 57 parameters $\eta$, $\eta_m$ and try to determine the
remaining ones by virtue of~(\ref{pure0}).
Equation $(\mathbf{133}^{-2})$ directly determines $\xi_\alpha$
\bea
\xi_{\alpha} &=& - \frac{6}{\eta} \, (t_{\alpha})^{mn} \, \eta_m \, \eta_n
\;.
\label{solu1}
\eea
With $(\mathbf{56}^{-1})$, we find for $\tilde\eta_m$
\bea
\tilde\eta_m &=& \frac{\xi}{\eta} \, \eta_m  - \frac{24}{\eta^2} \, 
(t^{\alpha})_m{}^{n} \, (t_{\alpha})^{pq} \, \eta_n \, \eta_p \, \eta_q
\;.
\label{solu2}
\eea
Equation $(\mathbf{133}^{0})$ is then automatically satisfied. Its verification
requires the vanishing of the term quartic in $\eta_m$ which simply
follows from the absence of an adjoint representation in the totally
symmetric tensor product $({\bf 56}^{\otimes4})_{\rm sym}$.

Continuing with the $\bf{133}^{+2}$, we obtain after using
(\ref{solu1}), (\ref{solu2})
\bea
\tilde \eta \, \xi_{\alpha} &=&
-\frac{6}{\eta^2} \, (t_{\alpha})^{mn} \, \eta_{m} \, \eta_n \, \xi ^2
-\frac{96}{\eta^2} \, (t_{\alpha})^{mn} \, (t^{\beta})_{m}{}^{k} \, (t^{\gamma})_{n}{}^{l} \, \eta_k \, \eta_l \, \xi_{\beta} \, \xi_{\gamma} \;.
\label{step1}
\eea
The last term can be simplified by means of the identity~(\ref{identity2}).
Multiplying the latter with six $\eta_m$'s, we find
\bea
 (t_{\alpha})^{mn} \, (t^{\beta})_{m}{}^{k} \, (t^{\gamma})_{n}{}^{l} \, \eta_k \, \eta_l \, \xi_{\beta} \, \xi_{\gamma}
&=& 
 \frac{\eta}{48} \, \xi_{\beta} \, \xi^{\beta} \, \xi_{\alpha} 
\;.
\nonumber
\eea
Equation~(\ref{step1}) can be solved by setting
\bea
\tilde\eta  &=&  \frac{\xi ^2}{\eta} - \frac{2}{\eta} \, \xi_{\alpha} \, \xi^{\alpha}\;.
\label{solu3}
\eea

We have thus determined all unknown parameters
and verified the solution (\ref{sol3}).
It is straightforward to check, that the two singlets in (\ref{pure0})
are automatically satisfied with (\ref{solu1}), (\ref{solu2}), (\ref{solu3}).
Furthermore, the $\bf{56}^{+1}$ reduces to
\bea
0
&=& \frac{2}{\eta} \, \Big(\eta_m \, \xi^{\beta} \, \xi_{\beta} 
- 8 \, (t^{\alpha})_m{}^{n} \, (t^{\beta})_n{}^{k} \, \eta_k \, \xi_{\alpha} \, \xi_{\beta} \Big)\;,
\eea
which can be verified upon multiplying the identity~(\ref{identity1})
with five $\eta_m$'s.
Finally, the ${\bf 1539}^0$ reduces to
\bea
0 &=&
 -\frac{12}{\eta^2}\,\Big(
2\eta_{[n} (t^{\alpha})_{m]}{}^{p}  (t_{\alpha})^{qr}\,\eta_p\eta_q\eta_r
 +9  (t^{\alpha})_{m}{}^{k}  (t^{\beta})_{nk} (t_{\beta})^{pq} (t_{\alpha})^{rs}\,\eta_p\eta_q\eta_r\eta_s
 \Big)\nonumber\\[1ex]
&&{}\quad -~{\rm trace} 
 \;,
\eea
which is another consequence of~(\ref{identity1}).

We have thus verified, that the solution (\ref{solu1}), (\ref{solu2}), (\ref{solu3})
satisfies all constraint equations~(\ref{pure0}).
In principle, there are two more equations to verify
which transform in the ${\bf 912}^{\pm1}$.
However, with the given solution all constraint equations translate into
relations among a product of $\eta_m$'s
transforming in the fundamental ${\bf 56}$. Since there is no ${\bf 912}$ 
representation in the corresponding 
completely symmetrized tensor products of the fundamental ${\bf 56}$,
every such constraint is automatically satisfied.
This finishes the proof of~(\ref{sol3}).

\bigskip
\bigskip
\bigskip

{\small


\providecommand{\href}[2]{#2}\begingroup\raggedright\endgroup

}

\end{document}